\documentclass[preprint]{aastex}
\usepackage{epsfig}
\usepackage{color}
\definecolor{brown}{rgb}{0.6,0.4,0.2}
\definecolor{purple}{rgb}{0.5,0,0.5}

\slugcomment{Accepted for publication in the Astronomical Journal}

\shorttitle{X-ray Observations of CTB 1 and HB21} 
\shortauthors{Pannuti et al.}

\begin{document}


\title{Mixed-Morphology Supernova Remnants in X-rays: Isothermal
Plasma in HB21 and Probable Oxygen-Rich Ejecta in CTB 1}

\author{Thomas G. Pannuti\altaffilmark{1}}
\affil{{\it Spitzer} Science Center, California Institute of Technology,
MS 220-6, Pasadena, CA 91125; Current Address: Space Science Center, Department of Earth 
and Space Sciences, 235 Martindale Drive, Room 212E, Morehead State University, 
Morehead, KY 40351}
\email{t.pannuti@moreheadstate.edu}

\author{Jeonghee Rho}
\affil{{\it Spitzer} Science Center, California Institute of Technology,
MS 220-6, Pasadena, CA 91125; Current Address: SOFIA Science and Mission Operations/USRA,
NASA Ames Research Center, MS 211-3, Moffett Field, CA 94035 }
\email{jrho@sofia.usra.edu}

\author{Kazimierz J. Borkowski}
\affil{Department of Physics, North Carolina State University,
Box 8202, Raleigh, NC, 27695}
\email{kborkow@unity.ncsu.edu}

\and

\author{P. Brian Cameron}
\affil{Division of Physics, Mathematics and Astronomy, 105-24, California
Institute of Technology, Pasadena, CA 91125}
\email{pbc@astro.caltech.edu}

\altaffiltext{1}{Guest User, Canadian Astronomy Data Center, which is
operated by the Dominion Astrophysical Observatory for the National
Research Council of Canada's Herzberg Institute for Astrophysics.}

\begin{abstract}
We present an analysis of X-ray observations of the Galactic supernova  remnants
(SNRs) HB21 (G89.0$+$4.7) and CTB 1 (G116.9$+$0.2), two well-known 
members of the class of mixed-morphology (MM) SNRs.  Our analysis 
draws upon observations of both SNRs made with the Advanced Satellite 
for Cosmology and Astrophysics ({\it ASCA}): we have also used an 
archived {\it Chandra} observation of CTB 1 as part of this effort.
We find a marked contrast between the X-ray properties of HB21 and CTB
1: in the case of HB21, the extracted spectra of the northwest and southeast
regions of the X-ray emitting plasma associated with the SNR can be fit with a 
single thermal model with marginally
enhanced silicon and sulfur abundances. For both of these regions, the derived 
column density and temperature are $N$$_H$ $\sim$ 
0.3$\times$10$^{22}$ cm$^{-2}$ and $\it{kT}$ $\sim$ 0.7 keV, respectively. 
No significant spatial differences in temperature or elemental abundances 
between the two regions are detected and the X-ray-emitting plasma for 
both regions is close to ionization equilibrium.  Our 
{\it Chandra} spectral analysis of CTB 1 reveals that this source is likely an 
oxygen-rich SNR with enhanced abundances of oxygen and neon: this result is 
quite surprising for an evolved SNR like CTB 1. The high angular resolution
{\it Chandra} observation of CTB 1 reveals spectral variations across this
SNR: in particular, we have detected localized hard emission with an angular 
extent of $\sim 1'$. The extracted {\it ASCA} spectra for
both the southwest and northeastern regions of CTB 1 cannot be fit 
with a single thermal component and instead an additional component is
required to account for the presence of excess emission seen at higher energies.
Based on our fits to the extracted {\it ASCA} spectra, we derive a column 
density $N$$_H$ $\sim$ 0.6$\times$ 10$^{22}$ cm$^{-2}$ and a temperature 
for the soft thermal component of $\it{kT}$$_{soft}$$\sim$ 0.28 keV for
both regions. The hard emission from the southwest region may be modeled with 
either a thermal component with a temperature $\it{kT}$$_{hard}$ $\sim$ 3 keV 
or by a power law component with a photon index $\Gamma$ $\sim$ 2-3; for
the northeast region the hard emission may be modeled with a power law 
component with a photon index $\Gamma$ = 1.4. The detection of center-filled
ejecta-dominated X-ray emission from HB21 and CTB 1 as well as other MM SNRs 
suggests a new scenario for the origin of contrasting X-ray and radio morphologies of 
this class of sources.  Lastly, we have analyzed the properties of the discrete hard X-ray 
source 1WGA J0001.4$+$6229 which is seen in projection just inside the northeastern
shell of CTB 1. Our extracted {\it ASCA} GIS spectra of this source are best
fit using a power-law model with a photon index $\Gamma$=2.2$^{+0.5}_{-1.2}$:
this slope is typical for featureless power-law continua produced by
rotation-powered pulsars. This source may be
a neutron star associated with CTB 1. We find marginal evidence for X-ray
pulsations from this source with a period of 47.6154 milliseconds. A
deep radio observation of this source failed to reveal any
pulsations.
%Based on a timing analysis of the {\it ASCA} data for this source,
%we have detected (at a significance of $\sim$4$\sigma$) a 47.6154 ms
%period in the X-ray emission: a complementary radio observation failed 
%to detect pulsed emission from this source. This object may be a
%newly-discovered neutron star physically associated with CTB 1, though 
%additional observations are needed to confirm this association.  
%(THE ABSTRACT WILL BE REWRITTEN AFTER THE TEXT IS FINISHED).
%CHECK
%1) XMM dat: O line and hard source.
\end{abstract}

\keywords{supernova remnants: general -- 
supernova remnants: individual (\objectname{HB21 (G89.0$+$4.7), CTB 1
(G116.9$+$0.2), 1WGA J0001.4$+$6229}) -- X-rays: ISM}

\section{Introduction}
A new morphological class of supernova remnants (SNRs) known as the 
mixed-morphology (MM) SNRs
has been firmly established in the recent literature \citep{Rho98, Shelton99}. 
The defining characteristics of SNRs of this class include
a shell-like radio morphology combined with a centrally-filled X-ray 
morphology. X-ray observations of these SNRs made with the 
$\it{Roentgensatellit}$ ({\it ROSAT}), the Advanced Satellite for 
Cosmology and Astrophysics ({\it ASCA}),  {\it Chandra} and {\it XMM-Newton} 
have found that the central 
X-ray emission from these SNRs is not non-thermal emission
(as would be expected from a central plerion) but instead thermal from 
shock-heated swept-up interstellar material. Examples of well-known
MM SNRs include W28 \citep{Rho02}, G290.1$-$0.8 (MSH 11$-$6$\it{1}$A) 
\citep{Slane02} and IC 443 \citep{Kawasaki02}. \citet{Rho98} suggested 
that as many as 25\% of the entire population of Galactic SNRs may belong
to this morphological class. Based on CO and infrared observations 
\citep{Koo01, Reach05},
it appears that many MM SNRs are interacting with nearby molecular and HI
clouds. This result suggests a connection between the contrasting X-ray and 
radio morphologies of these SNRs with the interaction between these sources
and adjacent clouds, but such a connection 
lacks a detailed theoretical basis at this time. Two leading scenarios 
have been advanced to explain the origin of the contrasting radio and
X-ray morphologies of MM SNRs: in the first scenario -- known as the
evaporating clouds scenario -- molecular clouds overrun by the expanding
shock front of the SNR survive passage through the shock and eventually
evaporate, providing a source of material that increases the density of the
interior X-ray-emitting plasma of the SNR \citep{Cowie77, White91}. In
the second scenario -- known as the radiative shell model -- the SNR has
evolved to an advanced stage where the shock temperature is low and
very soft X-ray emission from the shell is absorbed by the interstellar
medium (ISM): therefore,
the only detectable X-ray emission is from the interior of the SNR
\citep{Cox99, Shelton99, Shelton04}. In the current paper we analyze and
discuss X-ray emission from two Galactic SNRs -- HB21 and CTB 1 -- which
have both been previously classified as MM SNRs by \citet{Rho98}. 
\par
HB21 (G89.0$+$4.7) was discovered in a radio survey by \citet{Brown53}.
The radio angular extent of this SNR is large -- 
120$\times$90 arcminutes \citep{Green09a} -- and the radio morphology is a
closed shell. The shell appears to be flattened 
along the eastern boundary and features bright regions along the northern and 
southern boundaries with a prominent indentation seen along the northern
boundary. The radio emission from HB21 is strongly polarized (3.7$\%$$\pm$0.4$\%$) with a 
projected magnetic field tangential to the shell \citep{Kundu71,Kundu73,Kothes06},
suggesting that the shell was compressed during the radiative evolutionary stage of the
SNR. The measured radio spectral index for this SNR is $\alpha$ $\sim$ 0.4 
($S_\nu \propto \nu^{-\alpha}$) \citep{Leahy06, Green09b} but significant variations
in the values (from 0.0-0.8 with a standard deviation of 0.16) of the spectral index across 
the face of the SNR were observed by \citet{Leahy06}. Based on {\it IRAS} observations,
\citet{Saken92} detected clumpy infrared filaments associated with this SNR. 
Filamentary optical emission from this SNR with an angular extent comparable to the
radio shell was detected by \citet{Mavromatakis07}. Extensive evidence exists that
indicates HB21 is interacting with adjacent molecular clouds: this evidence includes
CO observations \citep{Tatematsu90, Koo01, Byun06} as well as near- and mid-infrared 
observations of the sites of shock-molecular cloud interactions along the northern and
southern parts of the SNR \citep{Shinn09,Shinn10}.  HI 
observations toward HB21 \citep{Tatematsu90, Koo91} have revealed a high velocity 
expanding shell associated with this SNR. Finally, HB21 has
been the subject of prior pointed X-ray observations made by {\it Einstein}
\citep{Leahy87} and {\it ROSAT} \citep{Rho95, Leahy96}: we include in this 
paper
the {\it ROSAT} images presented previously by \citet{Rho95} in her
PhD thesis work. No pulsars or $\gamma$-ray sources are believed to be 
associated with HB21: radio searches for a pulsar associated with this SNR were
conducted by \citet{Biggs96} and \citet{Lorimer98} but no candidate sources were found. 
The distance to HB21 is not well known: \citet{Tatematsu90} argued for a distance of
only 0.8 kpc based on an association between the SNR and molecular material that 
belongs to the Cygnus OB7 association \citep{Humphreys78}.
However, \citet{Yoshita01} suggested a distance
of $\geq$1.6 kpc based on a correlation that those authors found between
X-ray absorbing column density and extinction, and \citet{Byun06}
suggested a distance of 1.7 kpc based on CO observations.
In this paper, we have adopted a distance of 1.7 kpc to HB21.
\par

%***************************************************************************
%IS RADIO EMISSION IN CTB 1 STRONGLY POLARIZED? IF SO, ONE NEEDS A REFERENCE.
%PRESENCE OF ABSENCE OF POLARIZATION IS IMPORTANT, BECAUSE PRESENCE OF 
%MAGNETIC FIELD TANGENTIAL TO THE SHELL MENAS THAT SHELL WAS
% COMPRESSED DURING
%THE RADIATIVE COOLING STAGE OF THE SNR.
%***************************************************************************
CTB 1 (G116.9$+$0.2) was discovered in a survey of Galactic radio emission 
at 960 MHz by \citet{Wilson60}. Subsequent radio observations of this SNR \citep{Velusamy74,
Angerhofer77,Landecker82, YarUyaniker04,Tian06,Kothes06} reveal a radio morphology 
that may be described as a nearly complete circular shell with a diameter of approximately 
34 arcminutes \citep{Green09b}. The
radio emission is brightest along the western rim and a prominent gap is seen along the
northern and northeastern sector of the circular emission. Like HB21, the magnetic field is 
aligned in the tangential direction, also suggesting that the shell was compressed during the 
radiative stage of the evolution of the SNR, but compared to HB21 the degree of polarization 
is much lower (0.4\%$\pm$0.1\% -- see \citet{Kothes06}). The measured spectral index 
of the observed radio emission is $\alpha$ $\sim$ 0.6 \citep{Landecker82, Kothes06, 
Tian06, Green09b}. CTB 1 has also been detected at 
optical wavelengths, in emission lines such 
as [\,OIII\,] $\lambda$5007 and [\,SII\,] $\lambda\lambda$ 6716, 6731; the 
observed optical shell-like morphology 
closely matches the radio shell. The optical images of CTB 1 presented by 
\citet{Fesen97} depict a remarkable contrast in the emission-line
properties of this SNR: while [\,SII\,] emission is seen from roughly the
entire optical shell (with the greatest amount of emission in the
south), the [\,OIII\,] emission appears to be almost entirely localized to
the western rim of the shell. \citet{Saken92} 
detected infrared emission from CTB 1 in the 60 $\mu$m and 100 $\mu$m 
 {\it IRAS} bands: an arc of infrared emission was seen in
the 60 $\mu$m/100 $\mu$m ratio image that appears to be coincident with
the radio shell. CTB 1 was observed in X-rays with {\it ROSAT} 
\citep{Hailey95,Rho95,Craig97}: like other MM SNRs, the X-ray emission from this SNR 
(which has a thermal origin) lies interior to the radio and optical shells. Remarkably, the
X-ray emission is also seen to extend through the known northern gap of the SNR.
Like HB21, no pulsars or $\gamma$-ray sources are believed to be 
associated with CTB 1: a radio search conducted by \citet{Lorimer98} for a pulsar
revealed no candidate sources. Published distance estimates
for this SNR have ranged from 1.6 to 3.5 kpc; in this paper we have adopted a 
distance to CTB 1 of 3.1$\pm$0.4 kpc as measured by \citet{Hailey94}.

\par

%In this paper we present properties of two
%MM SNRs, HB 21 (G89.0$+$4.7) and CTB 1 (G116.9$+$0.2) using data from
%{\it ASCA} observations of both SNRs and a {\it Chandra} observation of CTB 1. 
The organization of this paper is as follows: in
Section \ref{ObsandReduction} we describe the {\it ASCA} X-ray observations of HB21 and 
CTB 1 and the {\it Chandra} observations of CTB 1, including details of
data reduction. {\it ROSAT} and radio observations of these SNRs
are also described in this section. 
In Section 3 we present the results of our spectral analyses for
both HB21 and CTB 1 (in Section 3.1 and Section 3.2, respectively). 
In Section 4 we discuss the nature of the hard discrete X-ray source 1WGA
J0001.4$+$6229: we have discovered weak evidence for pulsed X-ray emission 
from this source
(which is seen in projection against CTB 1) and consider the possibility
that it is a neutron star associated with CTB 1. We also present a search for radio
pulsations from 1WGA J0001.4$+$6229. Interpretations of our X-ray results for HB21 and CTB 1 
are presented in Section 5 and Section 6, respectively.  
We also detected hard X-ray emission from CTB 1: we discuss the nature  of this emission 
in Section \ref{CTB1HardSection}.  Our preliminary results of this paper have been presented in
\citet{Pannuti04} after which we note that similiar data sets were analyzed and presented by
\citet{Lazendic06}. Our primary results of HB21 are in agreement with
and strengthen those of \citet{Lazendic06}; for CTB 1,
our paper presents extensive and thorough analysis of the {\it Chandra} and
{\it ASCA} data in
smaller-scale regions. We also report important new results for this SNR, including  
the probable detection of oxygen-rich ejecta from CTB 1 as well as spectral variations across
the object. In addition, an X-ray hard point-like source is identified and an analysis
of its X-ray and radio properties is presented. Finally, the 
conclusions of this work are summarized in Section 8.

\section{Observations and Data Reduction\label{ObsandReduction}}

\subsection{{\it ASCA} Observations of HB21 and CTB 1}
Because the X-ray emission from both HB21 and CTB 1 cover a large angular
extent on the sky, two pointed observations were made  
of each SNR with {\it ASCA} \citep{Tanaka94}, namely the southeast and northwest regions of
HB21 and the southwest and northeast regions of CTB 1 (see Table \ref{ASCAObsTable}
for details of these observations). These observations provided almost
a complete spatial coverage of the X-ray emitting gas in both SNRs. 
The data reduction was conducted
using the ``XSELECT" program (Version 2.2), which is available from the
High Energy Astrophysics Science Archive Research Center 
(HEASARC\footnote{see http://heasarc.gsfc.nasa.gov.}).  There were two 
types of instruments onboard {\it ASCA} -- the Gas Imaging Spectrometer 
(GIS) and the Solid-State Imaging Spectrometer (SIS) -- and both
of these instruments were composed of two units denoted as GIS2, GIS3,
SIS0 and SIS1, respectively. A single GIS unit sampled a field of view 
$\sim 50'$ in diameter and a background 
count rate of the GIS is 5 $\times$ 10$^{-4}$ counts cm$^{-2}$ sec$^{-1}$ 
keV$^{-1}$; in comparison, a single SIS unit sampled a field of view 
approximately $44' \times 44'$ in size. The 
nominal FWHM angular resolution of
both the GIS and SIS units were approximately 1 arcminute. 
The standard REV2 screening criteria were applied when reducing both
the raw GIS and SIS datasets. We used ``FMOSAIC" 
from the FTOOLS software package to generate an X-ray
map by combining the GIS2 and GIS3 maps. 
We used the FTOOL ``MKGISBGD" to prepare 
blank-sky background spectra and images for each extracted GIS source 
spectra: for these background datasets, point sources which are brighter than
approximately 10$^{-13}$ ergs cm$^{-2}$ sec$^{-1}$ have been removed. 
Similarly, background spectra were generated using standard SIS blank-sky 
datasets for analyzing the extracted SIS source spectra. 
The standard GIS2 and GIS3 response matrix files (RMFs) were used for
analyzing extracted the GIS source spectra while the FTOOL ``sisrmg" was
used to prepare RMFs for the extracted SIS source spectra. Finally, the 
FTOOL ``ASCAARF" was used to prepare ancillary response files (ARFs) for 
each extracted GIS and SIS source spectra. 

\subsection{{\it Chandra} Observation of CTB 1}

We have also analyzed an archival dataset from an additional X-ray 
observation of CTB 1 made with {\it Chandra} \citep{Weisskopf02}.
The corresponding ObsID of this observation is 2810 (PI: S. Kulkarni)
and it was conducted as part of a search for
central X-ray sources associated with Galactic SNRs. This observation
was conducted in FAINT Mode on 14 September 2002 with the Advanced
CCD Imaging Spectrometer (ACIS) at a focal plane temperature of 
$-$120$^{\circ}$C such that the ACIS-I array of chips sampled a significant
portion of the X-ray emitting plasma located
interior to the radio shell of the SNR. The ACIS-I array is composed
of four front-illuminated CCD chips: each chip is 8$\farcm$3 $\times$ 
8$\farcm$3 and the field of view of the entire array is 
approximately 17$'$ $\times$ 17$'$. 
These chips are nominally sensitive to photons in
the 0.2 through 10 keV energy range: the maximum effective collecting 
area for each chip is approximately 525 cm$^{2}$ at 1.5 keV. The full-width at half-maximum
(FWHM) angular resolution of each chip at 1 keV is 1$"$ and finally the spectral resolution
at 1 keV of each chip is 56.
These data were reduced using the {\it Chandra} Interactive Analysis of
Observations (CIAO\footnote{http://cxc.harvard.edu/ciao/}) package
(Version 3.1) with the calibration database (CALDB) version 2.29. 
Standard processing was applied to this dataset: in particular, the task
``acis$\_$process$\_$events" was used to generate a new event file where
corrections for charge transfer inefficiency and time-dependent gain 
were applied. The data were also filtered for bad pixels, background
flare activity and events which had a GRADE value of 1, 5 or 7. 
Finally, we applied the good time interval (GTI) file supplied by
the pipeline (as well as the GTI file prepared when filtering for
background flares) and the resulting total effective exposure time 
of the observation was 48.9 kiloseconds. 
\par
Discrete sources were identified with the {\it CIAO} 
wavelet detection routine ``wavdetect" \citep{Freeman02}: in making a
final image, these sources were excluded and the image was exposure-corrected
and smoothed with the {\it CIAO} task ``csmooth."  We extracted spectra from several regions 
of the diffuse X-ray emission seen in the {\it Chandra} images using the {\it CIAO} task
``dmextract": results of the spectral analysis are presented in Section 
\ref{CTB1SubSection}. When extracting spectra, we 
excluded point sources identified by ``wavdetect" to help reduce
confusion with emission from background sources. We prepared ARFs and RMFs using
the CIAO tools ``mkwarf" and ``mkrmf," respectively; background spectra
were generated using a reprojected blank sky observation made with the
ACIS-I array and available from the {\it Chandra} X-ray Center via the
World Wide Web.\footnote{See http://cxc.harvard.edu/contrib/maxim/acisbg/.}
%Finally, each spectrum was grouped using the FTOOL ``grppha" such that there 
%were a minimum of 25 counts per channel. We comment that the apparent extent 
%of the X-ray emitting plasma associated with CTB 1 is far larger than the
%nominal 16$\arcmin$$\times$16$\arcmin$ area sampled by the ACIS-I array
%and that additional observations of different regions of this X-ray emitting
%plasma are required to evaluate more thoroughly the X-ray properties of this
%plasma.

\subsection{Additional Observations}

We also included {\it ROSAT} Position Sensitive Proportional Counter (PSPC)
observations of HB21 in our analysis: these observations were discussed
already in some detail in the Ph.D thesis of \citet{Rho95}. They extend
beyond the two fields observed by {\it ASCA} and provide the complete
spatial coverage of the SNR. The PSPC images are exposure and
particle background corrected and merged together using the analysis
techniques for extended objects developed by \citet{Snowden94}. The 
smoothing technique includes neighboring pixels within a
circle of increasing radius until a selected number of counts is
reached to optimize the signal-to-noise ratio.

Lastly, we have augmented the X-ray datasets analyzed in this work
with radio data provided by the Canadian Galactic Plane Survey
(CGPS) \citep{Taylor03}. From this survey
we have obtained radio images of HB21 and CTB 1 at the frequencies of
408 MHz and 1420 MHz. The angular resolution and sensitivity of
the 408 MHz data are 3.$'$4 $\times$ 3.$'$4 csc $\delta$ and 0.75 sin
$\delta$ K (3.0 mJy beam$^{-1}$), respectively, while the angular resolution 
and sensitivity of the 1420 MHz data are 1' $\times$ 1' csc $\delta$ and 71 
sin $\delta$ K (0.3 mJy beam$^{-1}$), respectively. The reader is referred to 
\citet{Taylor03} for more description about the CGPS radio observations 
and the accompanying data reduction process. 

\section{Results}

\subsection{HB21\label{HB21Section}}

\begin{figure}
\epsscale{1.0}
\plotone{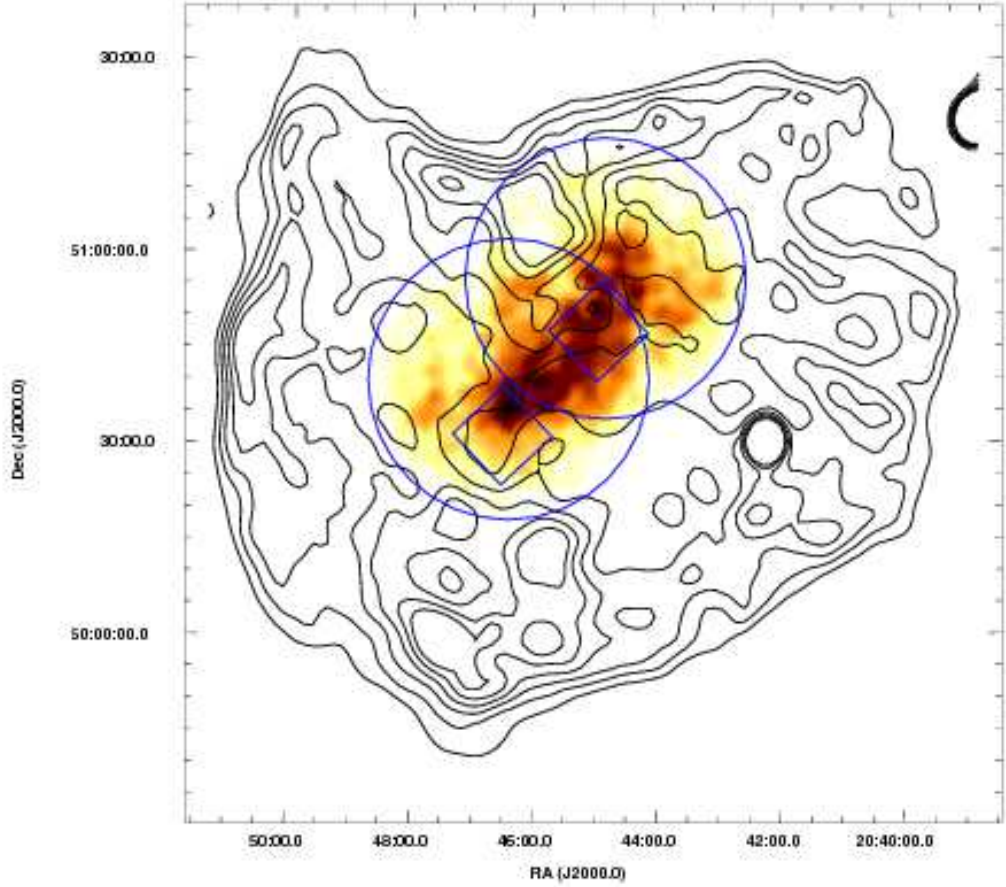}
\caption{
Mosaicked broadband {\it ASCA} GIS image of HB21: the emission
has been smoothed with a 1$\arcmin$ FWHM Gaussian. The intensity range 
is 0 -- 7.39$\times$10$^{-5}$ counts sec$^{-1}$ arcmin$^{-2}$. The
contours represent radio emission from HB21 as observed at 408 MHz with the 
Canadian Galactic Plane Survey; they range from
0.445 Jy to 0.890 Jy per 3$\farcm$6 $\times$ 3$\farcm$6) in steps of 0.0445 
Jy/beam.
% these contours range from T$_B$=100 K
%through 200 K in steps of 10 K. 
The blue boxes and 
circles represent the approximate fields of view of the {\it ASCA} SIS 
and the GIS, respectively.}
\label{hb21asca}
\end{figure}

\begin{figure}
\epsscale{1.2}
\plotone{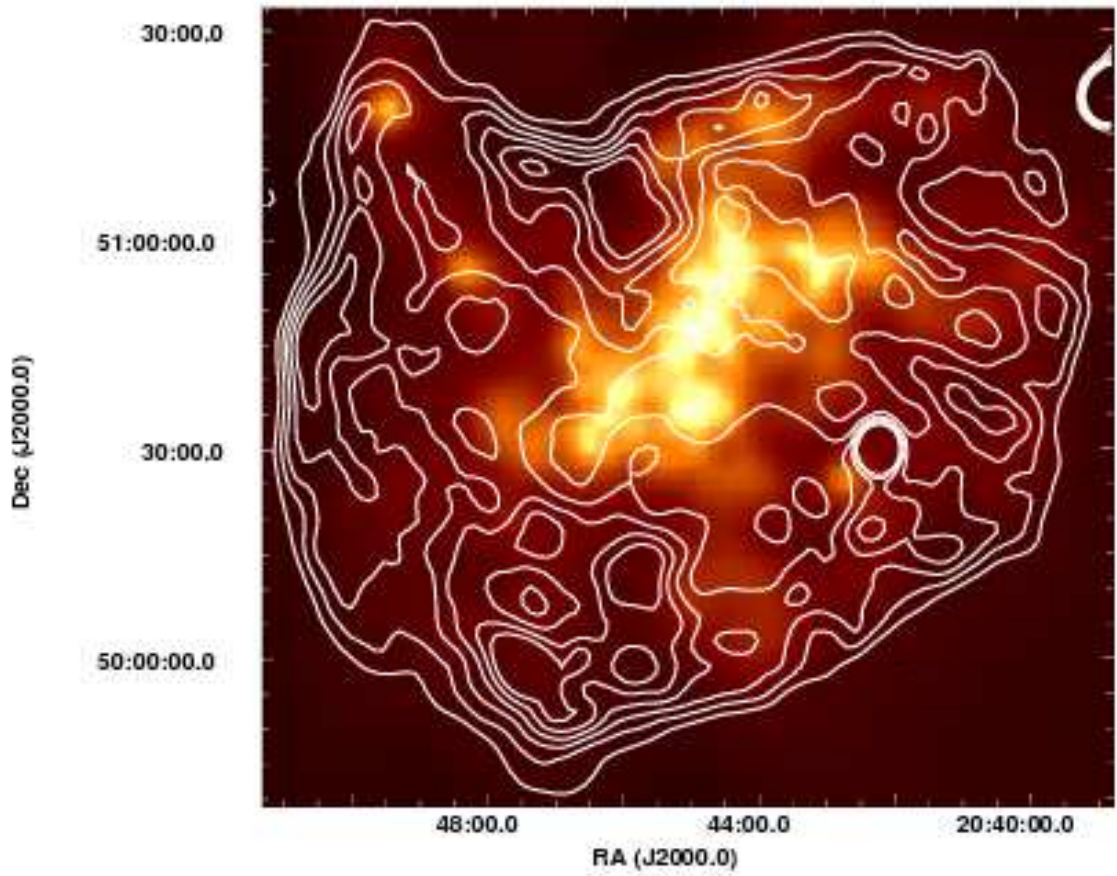}
\caption{Mosaicked {\it ROSAT} PSPC image of HB21: we have superposed the
same radio contours as in Figure \ref{hb21asca}. The intensity range of this
image is 10$^{-4}$ -- 2.1$\times$10$^{-2}$ counts sec$^{-1}$ arcmin$^{-2}$.}
\label{hb21rosat}
\end{figure}

\begin{figure}
\epsscale{1.2}
%{\hbox{
\psfig{figure=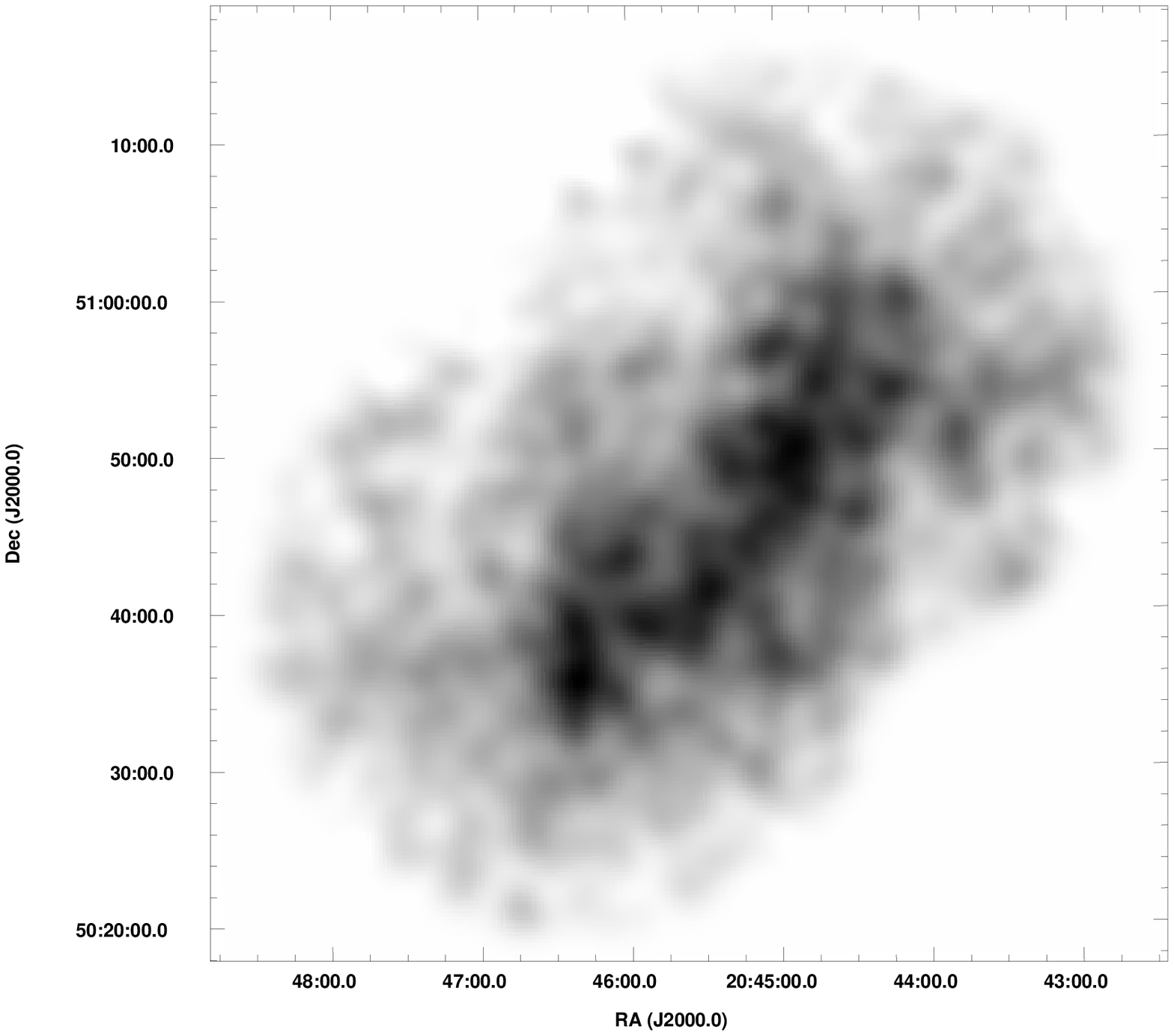,width=9truecm}
\psfig{figure=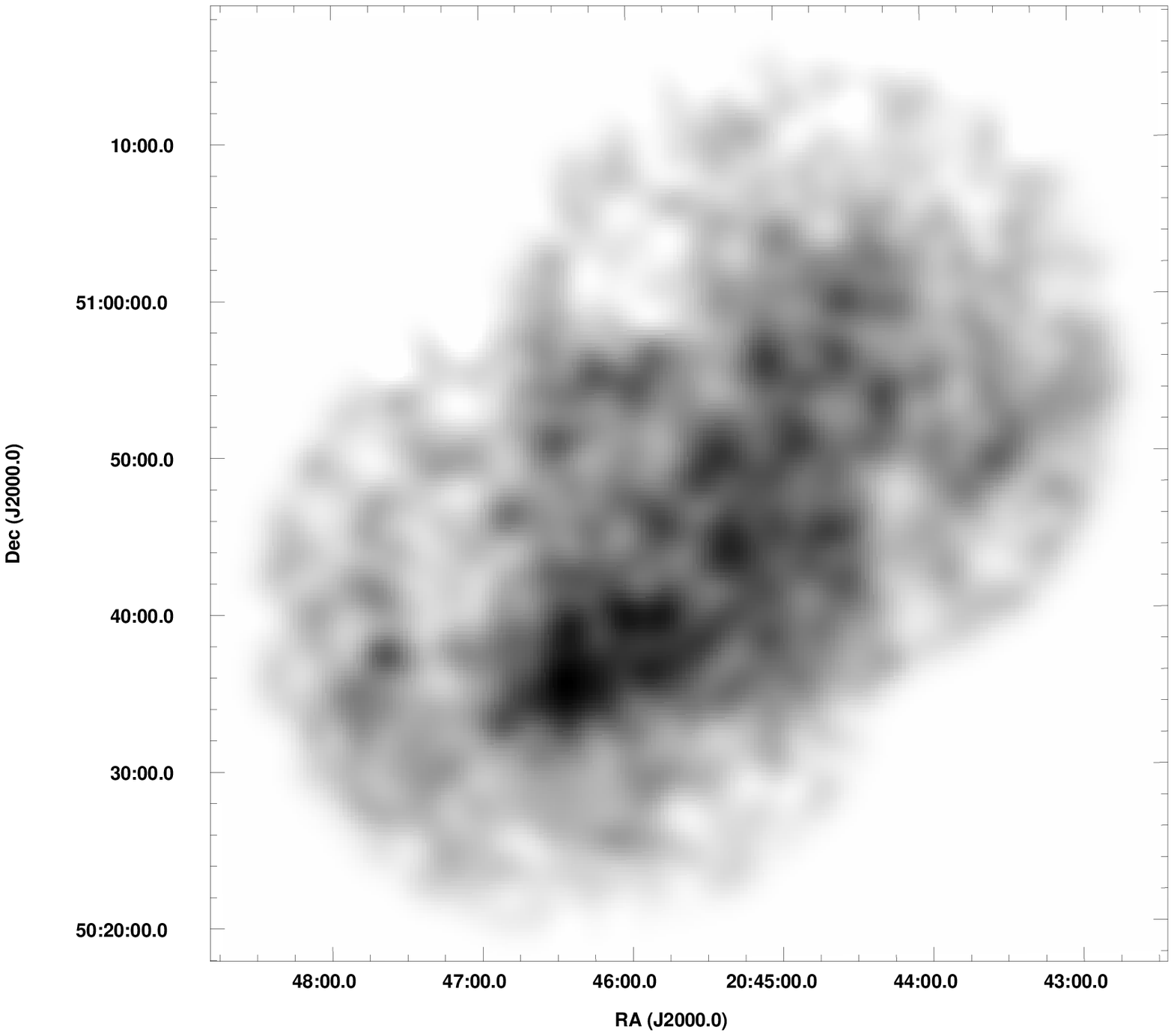,width=9truecm}
%}}
\caption{Mosaicked soft band (E$<$1 keV -- top) and hard band (E$>$1 
keV -- bottom) GIS images of the central region of HB21: both images have been smoothed with
a 1$\arcmin$ FWHM Gaussian. The coverages of the GIS images are marked in Figure \ref{hb21asca}.   The intensity ranges of the two images are 0 --
1.02$\times$10$^{-4}$ and 0 -- 2.58$\times$10$^{-5}$ counts sec$^{-1}$
arcmin$^{-2}$, respectively.}
\label{hb21softhard}
\end{figure}

In Figure \ref{hb21asca} we present our broadband (0.7-10.0 keV) 
exposure-corrected mosaicked {\it ASCA} GIS image of HB21: we have
overlaid radio emission contours using CGPS observations at a 
frequency of 408 MHz to show the extent of the SNR radio shell.
In Figure \ref{hb21rosat} we present
a mosaicked {\it ROSAT} PSPC image of HB21 (with the same radio
contours overlaid) which depicts the entire extent of X-ray emission 
from the SNR (which extends beyond the two fields observed by {\it ASCA}).
It is clear from inspection of these images that the X-ray
emission is located in the interior of the well-defined SNR radio shell:
this combination of X-ray and radio morphologies exemplifies the
defining characteristics of mixed-morphology SNRs. The bulk of the
interior X-ray emitting plasma is located just south of a prominent
bend in the northern edge of the radio shell. We note that
\citet{Koo01} detected broad CO emission lines from the location of
this bend and \citet{Shinn09} presented near- and mid-infrared images of this
same region which showed shock-cloud interaction features. Both of these
studies indicated that this is a site of an interaction between
HB21 and a neighboring molecular cloud complex.  In Figure
\ref{hb21softhard} we present additional mosaicked {\it ASCA} GIS
images which depict soft and hard emission (corresponding to the energy
ranges of E$<$1 keV and E$>$1 keV, respectively) from this SNR.  
The GIS2 count rates for E$<$1 keV  and E$>$1 keV are
1.91($\pm$0.08)$\times$10$^{-2}$  cts s$^{-1}$ and
3.21($\pm$0.15)$\times$10$^{-2}$ cts s$^{-1}$, respectively, while the
respective count rates  for  E$<$2 keV  and E$>$2 keV are
4.56($\pm$0.13)$\times$10$^{-2}$ cts s$^{-1}$ and
0.56($\pm$0.08)$\times$10$^{-2}$ cts s$^{-1}$. Significantly more X-ray
emission is detected from HB21 at E$<$2 keV X-ray energies than at E$>$2 keV
energies, illustrating the soft spectral nature typical of SNRs.
\par
The GIS2/GIS3 spectra were extracted from elliptical regions approximately 
23$\arcmin$ in size carefully selected to include most of the emission and 
avoid the edges of the field of view. Likewise, the SIS0/SIS1
spectra were extracted from square-shaped regions approximately 10
arcminutes on a side and again the edges of the fields of view were
avoided. The spectral extraction regions are marked in Figure 
\ref{hb21asca}. We analyzed the extracted spectra using the software
package {\it XSPEC}\footnote{http://heasarc.gsfc.nasa.gov.docs/xanadu/xspec/.}
Version 11.3.1 \citep{Arnaud96}. 
For our spectral fitting we used two thermal models: the 
thermal model VAPEC, which describes an emission spectrum from a 
collisionally-ionized diffuse gas with variable elemental abundances 
\citep{Smith00, Smith01a, Smith01b}\footnote{Also see 
http://hea-www.harvard.edu/APEC.} and VNEI, which is 
a non-equilibrium collisional plasma model which assumes a constant 
temperature and single ionization parameter \citep{Hamilton83, 
Liedahl95, Borkowski01}. Photoelectric absorption along the line of sight was 
accounted for with the PHABS model; finally, we allowed the abundances 
of silicon and sulfur to vary during the fitting process 
(because lines associated with these particular elements are
noticeable in the spectra) while
leaving the abundances of the other elements frozen to solar values.
%After background subtraction, the count rates for the GIS2, GIS3, SIS0 
%and SIS1 observations of the northwestern portion of HB21 were 
%0.074$\pm$0.002, 0.073$\pm$0.002, 0.068$\pm$0.002 and 0.055$\pm$0.002
%counts per second over the 0.6-10.0 keV energy range, respectively. Likewise, 
%after background subtraction, the count rates for 
%the GIS2, GIS3, SIS0 and SIS1 
%bservations of the southeastern portion of this SNR over the same energy 
%range were 0.077$\pm$0.002, 0.079$\pm$0.002, 0.048$\pm$0.002 and 
%0.032$\pm$0.002 counts per second, respectively. 
\par
\begin{figure}
%\epsscale{1.0}
%\plotone{f4.ps}{rotation=90}
\epsfig{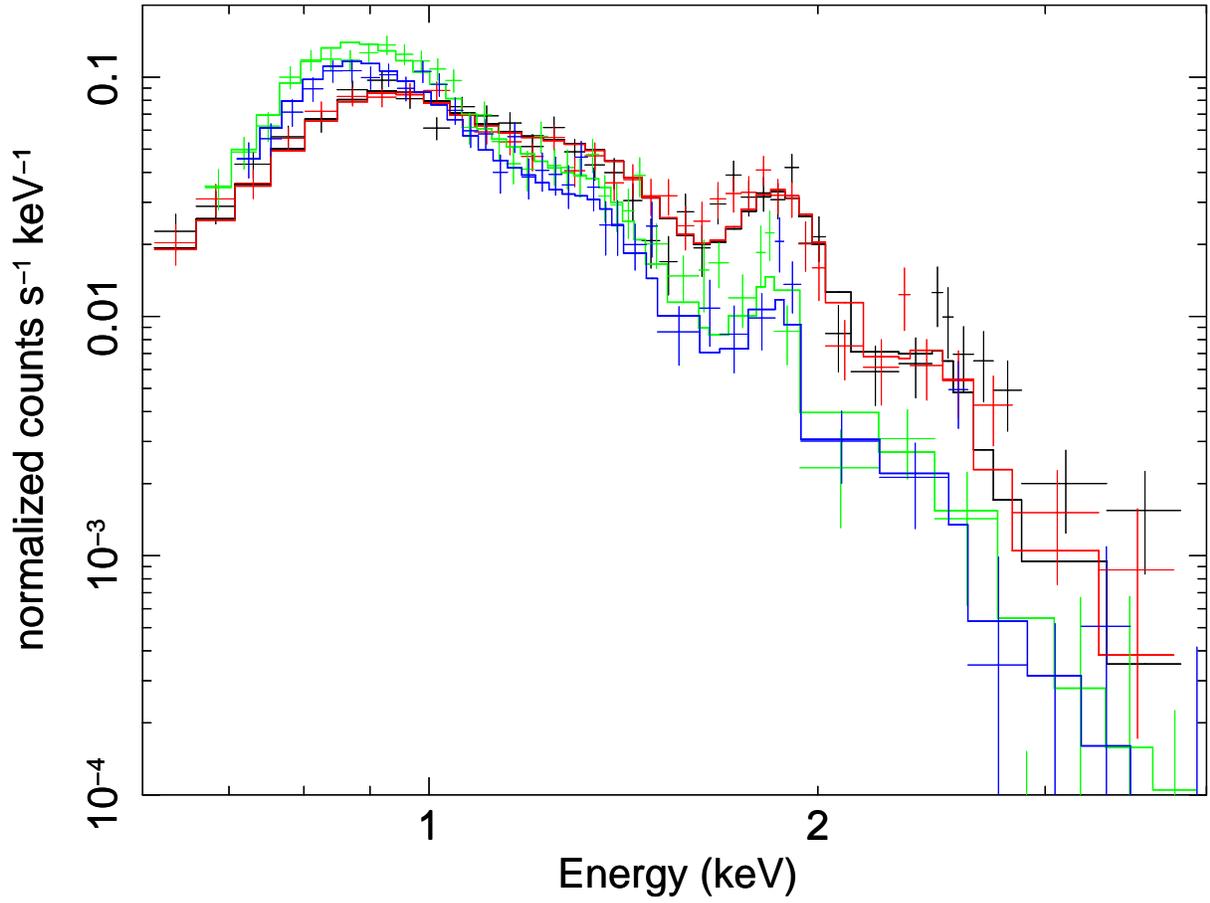}
\caption{GIS2 (red), GIS3 (black), SIS0 (green) and SIS1 (blue) spectra of the 
northwestern portion of
HB21 superposed on PHABS$\times$VNEI models (see Table \ref{HB21SpectralTable}).}
\label{hb21spectra}
\end{figure}

In Table \ref{HB21SpectralTable} we present results of our simultaneous
fits to the GIS2/GIS3 and SIS0/SIS1 spectra for both the northwestern
and southeastern regions of HB21. We have obtained 
statistically acceptable fits (with $\chi$$^2_{\nu}$ values of $\sim$ 1.04-1.06) 
to the extracted spectra using 
a single thermal component (that is, either the VAPEC model or the
VNEI model) for both the northwestern and southeastern
regions. The column density and temperature are 
similar for both thermal models, namely $N$$_H$ 
$\sim$ 2-3$\times$10$^{21}$ cm$^{-2}$ and $\it{kT}$ $\sim$ 0.66-0.68 keV.
In Figure \ref{hb21spectra}, we present the extracted GIS2, GIS3, SIS0 and SIS1
spectra for the northwest region of HB21: in each case, the fits obtained 
using ionization equilibrium (CIE) and nonequilibrium ionization models 
are comparable in quality. In addition, the abundances of silicon 
and sulfur in our spectral fits exceed solar abundances: these {\it ASCA} observations
are the first to reveal enhanced abundances of heavy elements in the X-ray spectra of HB21
(as also noticed by \citet{Lazendic06}). Our results are consistent with previous analyses of 
X-ray emission from HB21 \citep{Leahy87, Rho95} where our present analysis of the broadband 
{\it ASCA} spectra has yielded similar temperatures to those derived from {\it Einstein}
and {\it ROSAT} observations. However, with the data from the {\it ASCA} observations we 
may establish more stringent constraints on the
ionization timescale and the abundances of sulfur and silicon.  
The ionization timescales derived with the VNEI model for the two regions 
are long ($\tau$ = 5.9 ($>$3.2)$\times$10$^{11}$ cm$^{-3}$ s 
and $\tau$ = 4.1$^{+5.9}_{-1.1}$$\times$10$^{11}$ cm$^{-3}$ s for the northwestern and
southeastern regions of HB21, respectively) and within the error bounds for this
parameter, CIE is included ($\tau$ $\geq$ 10$^{12}$ cm$^{-3}$ s -- see \citet{Smith10}). 
For this reason and because both the VAPEC and the
VNEI models return equally acceptable fits, we argue that the X-ray emitting 
plasma located within the interior of HB21 is close to ionization equilibrium.
Similar to the results presented by \citet{Lazendic06},
we also observed slightly enhanced silicon and sulfur abundances in HB21: based on our 
PHABS$\times$VAPEC (PHABS$\times$VNEI) fits, our measured abundances are
Si=1.3$^{+0.3}_{-0.2}$ (1.8$\pm$0.05) and S=2.4$\pm$1.0 (3.6$^{+1.7}_{-1.9}$)
%%(90\% confidence levels) 
for the northwestern region and Si=1.4$\pm$0.3 (2.0$\pm$0.4) and S=1.7$^{+0.9}_{-0.8}$
(3.0$\pm$1.4) for the southeastern region (see Table \ref{HB21SpectralTable}). 
The slightly different abundances between the VAPEC and VNEI fits may be due 
to either differences in the predicted line strengths between the non-equilibrium and equilibrium
conditions or different sets of atomic 
data incorporated in these models. The respective lower limits for the abundances of Si (S) for 
the northwestern region are 1.1 (1.3)  and 1.4 (1.7), while the lower limits for the 
southeastern region are 1.1 
(1.6) and 0.9 (1.6). The northwestern region shows modestly more consistent evidence of 
slightly enhanced silicon and sulfur abundances than the southeastern region.
%at the 99\% confidence level, Si=1.3$^{+0.5}_{-0.3}$ and 
%S=2.4$\pm$1.4 for the northwest and Si=1.4$\pm$0.4 and S=1.7$^{+1.3}_{-1.2}$, 
%for the southeast, respectively. 
This suggests a contribution to the observed X-ray spectra from ejecta material. 
\par
\begin{figure}
\epsscale{1.0}
\psfig{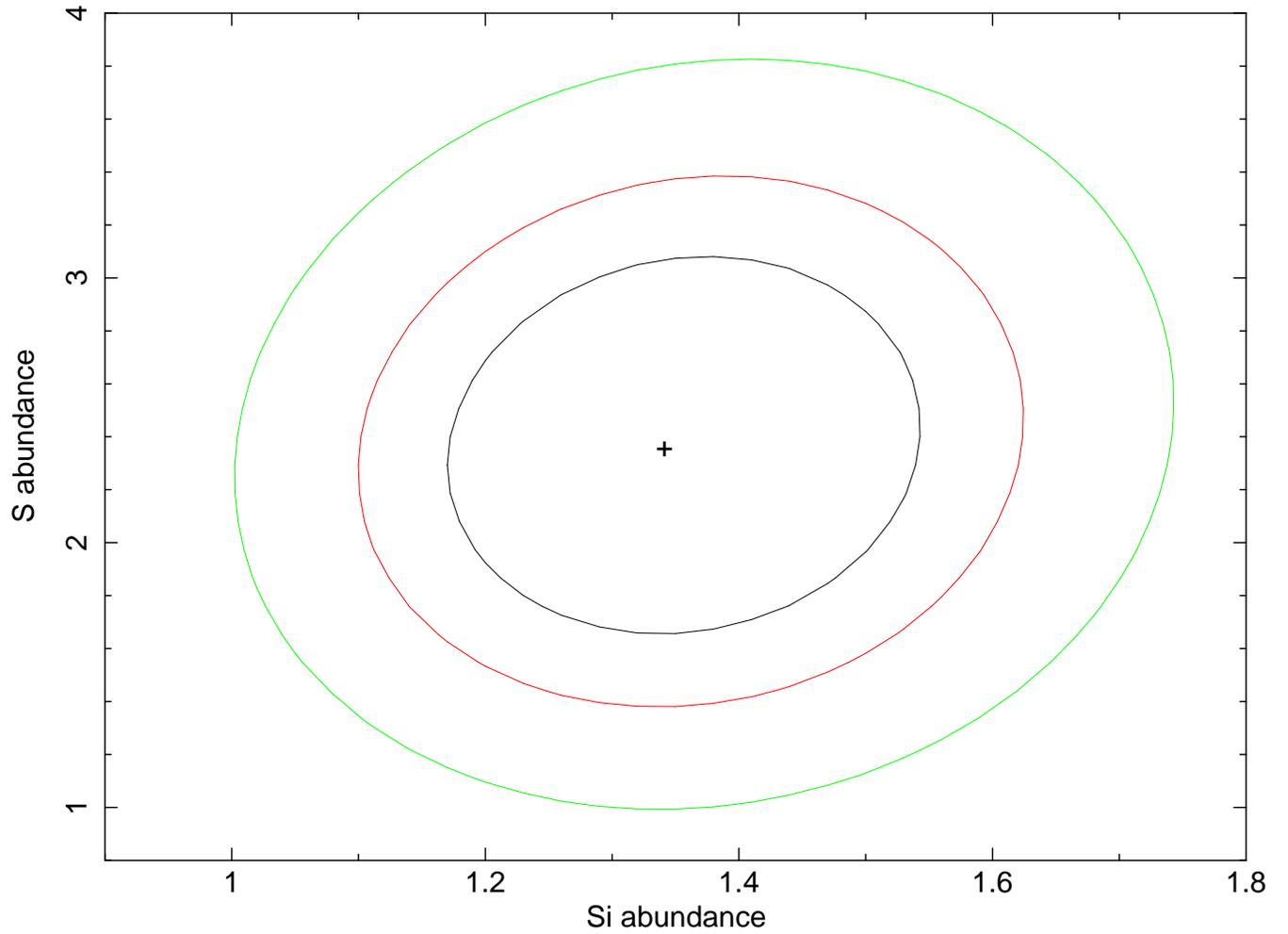}
\caption{Confidence contours for the abundances of Si and S for the PHABS$\times$VAPEC
fit to the extracted GIS2, GIS3, SIS0 and SIS1 spectra of the northwestern portion of HB21.
The confidence contours are at the 1$\sigma$, 2$\sigma$ and 3$\sigma$ levels (see Table
\ref{HB21SpectralTable}).}
\label{HB21ConfContours}
\end{figure}

In Figure \ref{HB21ConfContours} we present a plot of confidence contours for the
silicon and sulfur abundances based on the PHABS$\times$VAPEC fit to the spectrum of the
northwestern region. No significant spectral differences are seen between the
northwestern and southeastern regions of this SNR. In their analysis
of spectra extracted from the {\it ASCA} observations of HB21, \citet{Lazendic06} 
found that fits obtained using the VNEI model to the spectra extracted for both regions 
gave better fits at a statistically-significant level (4$\sigma$) than fits obtained by a thermal
plasma. In contrast, we find that fits obtained with VNEI and fits obtained with a standard 
thermal plasma are comparable in quality. Also, \citet{Lazendic06} derived comparable
(though slightly larger) values for $N$$_H$ for both regions: those authors also found a
similar trend where $N$$_H$ is modestly elevated for the southeastern region compared
to the northwestern region. We suspect that the minor differences between our
results and those obtained by \citet{Lazendic06} may be attributed to small differences in 
data reduction and spectral
analysis techniques (such as background subtraction). We also point out that 
\citet{Lazendic06} commented that the derived column density
values seemed to be rather high if HB21 is indeed only 0.8 kpc distant (which was their
assumed distance to the SNR). They noted that an
elevated column density (at least for the eastern portion of the SNR) is comparable
to the column density of the complex of molecular clouds seen toward HB21 as described
by \citet{Tatematsu90}. We point out that if the larger distance to HB21 that we have assumed
(1.7 kpc) is adopted, the measured values for $N$$_H$ seem more reasonable.
\par
Lastly, to help determine more stringent constraints on the properties of
the X-ray emitting plasma (specifically the abundances of silicon and sulfur), we 
simultaneously fit the GIS2 spectra extracted for the northwestern and southeastern portions 
of the SNR with first the PHABS$\times$VAPEC model and then with the 
PHABS$\times$VNEI model. For both models we obtained statistically acceptable fits 
(with $\chi$$^2$$_{\nu}$ values of $\sim$ 1.1) with values for $kT$ and $N$$_H$ that
were consistent with those obtained for fits of the individual regions (that is, $kT$ $\sim$
0.6 keV and $N$$_H$ $\sim$ 0.2 $\times$ 10$^{22}$ cm$^{-2}$). The abundances for
both silicon and sulfur were indeed enhanced relative to solar: for the PHABS$\times$VAPEC 
(PHABS$\times$VNEI) model, the abundances were 1.7$\pm$0.4 (1.9$^{+0.5}_{-0.4}$) for
silicon and 3.5$^{+1.9}_{-1.0}$ (4.3$^{+2.2}_{-1.8}$) for sulfur. The ionization timescale
derived for the PHABS$\times$VNEI model -- $\tau$ $\sim$ 4($>$0.03)$\times$10$^{13}$
cm$^{-3}$ s -- is also consistent with ionization equilibrium.We summarize the results
for these fits in Table \ref{HB21SpectralTable}.

\subsection{CTB 1\label{CTB1SubSection}}

\begin{figure}
\epsscale{1.0}
\psfig{figure=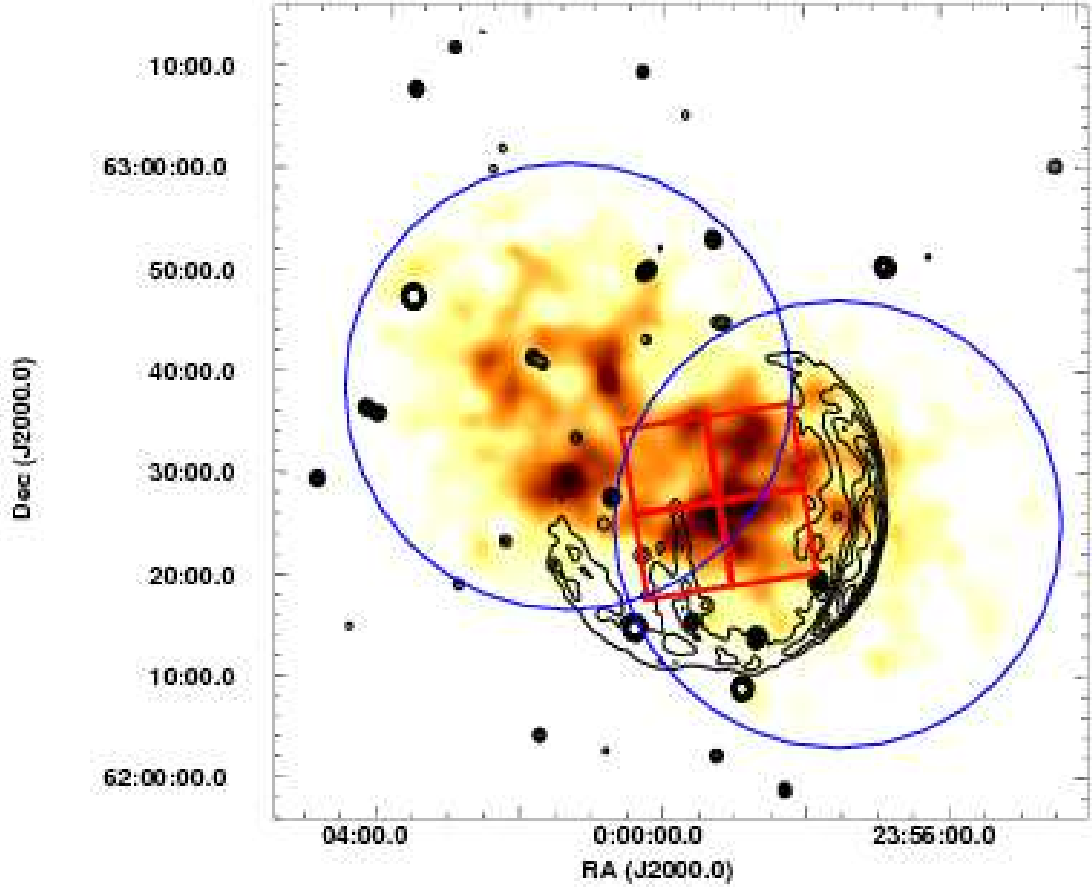,height=15truecm}
\caption{Mosaicked broadband {\it ASCA} GIS image of CTB 1 (color): the 
emission has been smoothed with a 1$\arcmin$ FWHM Gaussian. The intensity
range is 0 -- 5.93$\times$10$^{-5}$ counts sec$^{-1}$ arcmin$^{-2}$.
The contours represent radio emission as observed at 408 MHz with the 
Canadian Galactic Plane Survey (CGPS):  the contours range
from 0.024 Jy to 0.063 Jy per 0$\farcm$9 $\times$ 0$\farcm$9 in steps of 
0.0035 Jy/beam.
%these contours range from T$_B$=7 K
%through 18 K in steps of 1 K. 
The blue circles depict the approximate fields of view of the {\it ASCA} GISs 
while the red squares depict the approximate fields of view of the 
{\it Chandra} ACIS-I chips.}
\label{ctb1asca}
\end{figure}

\begin{figure}
\psfig{figure=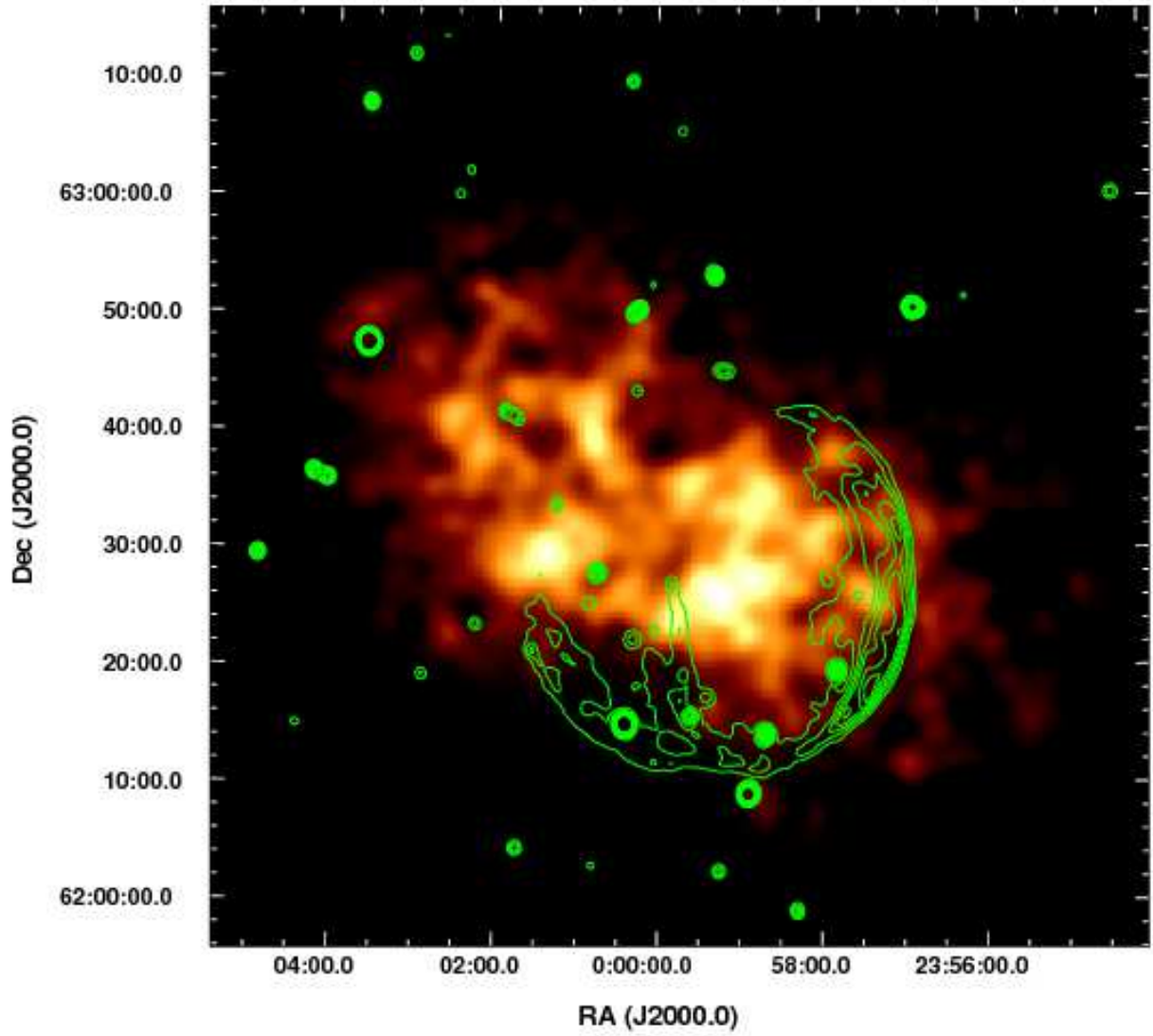,height=15truecm}
\caption{{\it ASCA} GIS image of CTB 1 with CGPS radio contours overlaid: these radio
contours are at the same levels as those shown in Figure \ref{ctb1asca}.}
\label{ctb1asca2}
\end{figure}

\begin{figure}
\epsscale{1.2}
%{\hbox{
%\psfig{figure=CTB1_LT1_MOSAIC.ps,width=9truecm}
%\psfig{figure=CTB1_GT1_MOSAIC.ps,width=9truecm}
%}}
\plottwo{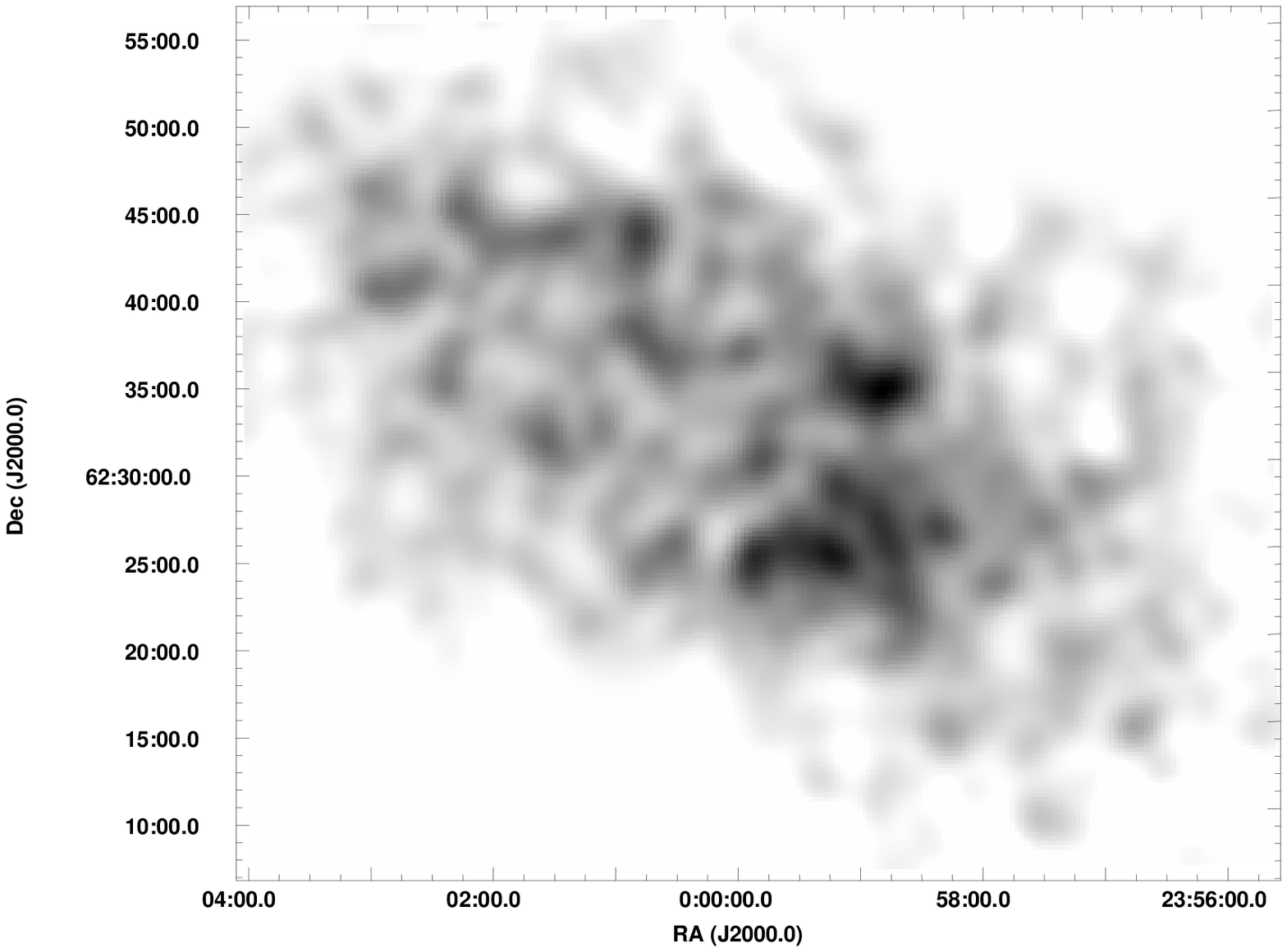}{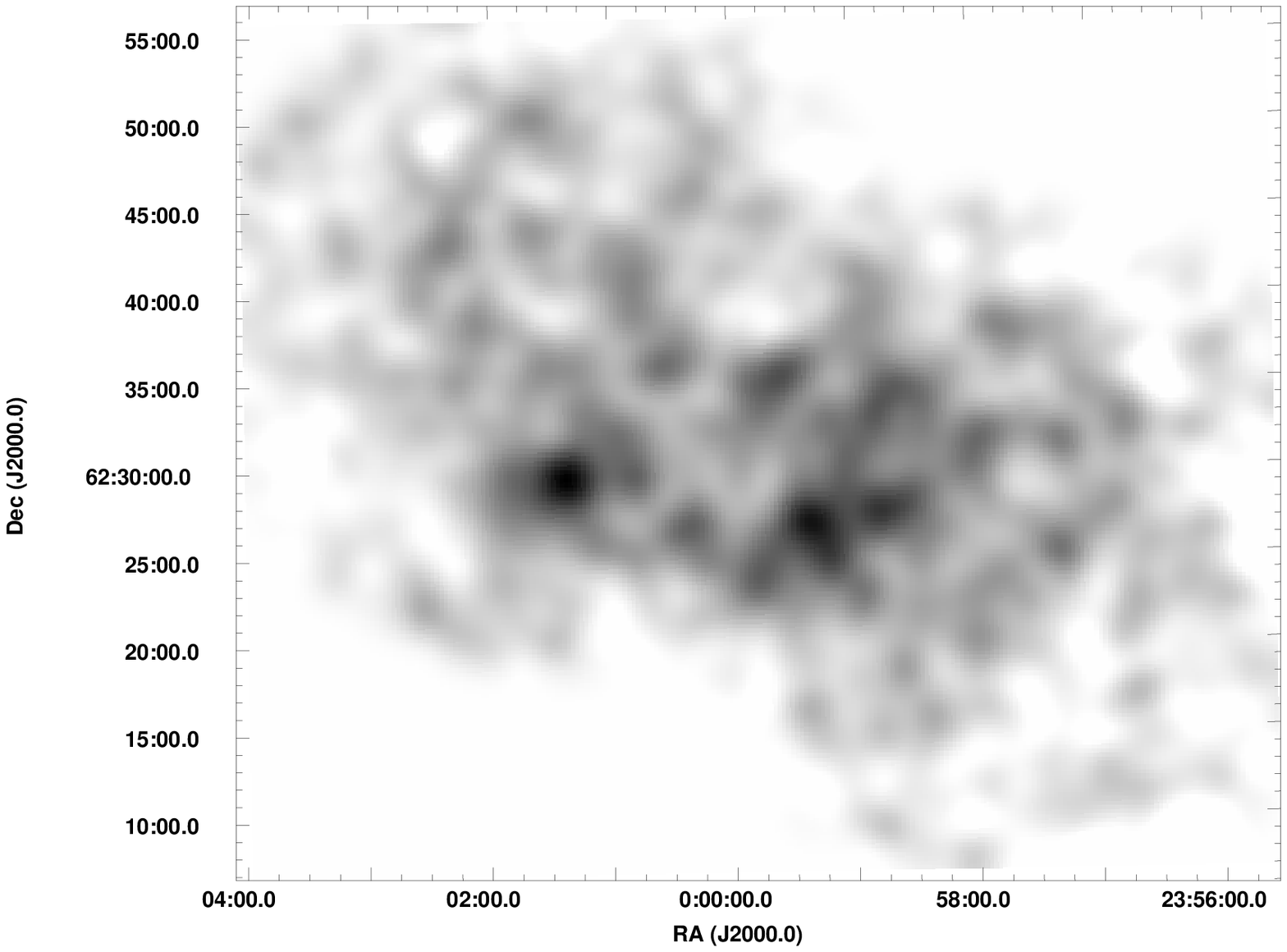}
\caption{Mosaicked soft band (E$<$1 keV -- top) and hard band (E$>$1 
keV -- bottom) GIS images of CTB 1: both images have been smoothed with
a 1$\arcmin$ FWHM Gaussian. The intensity ranges are
0 -- 6.42$\times$10$^{-5}$ and 0 -- 2.85$\times$10$^{-5}$ counts sec$^{-1}$ 
arcmin$^{-2}$, respectively. The discrete source 1WGA J0001.4$+$6229 
(located at RA (J2000.0) 00$^h$ 01$^m$ 25$^s$.5, Dec (J2000.0) $+$62$^{\circ}$ 
29$\arcmin$ 40$\arcsec$) becomes more prominent at higher energies: this 
source is discussed in detail in Section \ref{WGASection}.}
\label{ctb1softhard}
\end{figure}

\begin{figure}
\epsscale{1.0}
\plotone{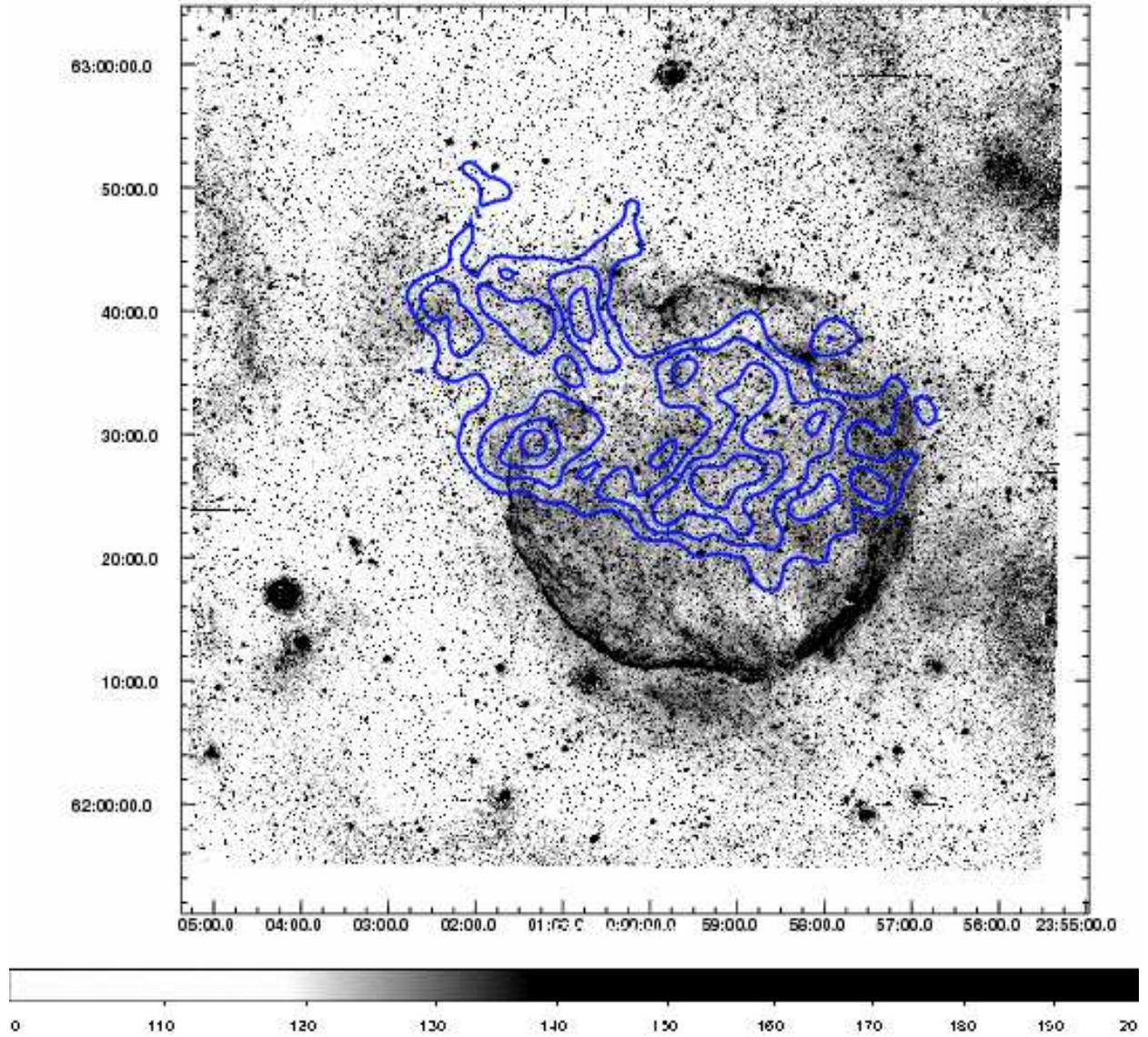}
\caption{H$\alpha$ image of CTB 1 in greyscale (courtesy of Robert Fesen)
with blue {\it ASCA} GIS contours overlaid. The contour levels are
2.05, 3.04, 4.02 and 5.01 $\times$ 10$^{-5}$ counts sec$^{-1}$
arcmin$^{-2}$. Notice how the X-ray emission extends through the
prominent gap of optical emission.}
%\caption{H$\alpha$ image of CTB 1 with broadband {\it ASCA} contours
%overlaid.
\label{ctb1optical}
\end{figure}

A broadband (0.7--10.0 keV) exposure-corrected and mosaicked {\it ASCA} GIS
image of CTB 1 is presented in both Figures \ref{ctb1asca} and \ref{ctb1asca2}
with radio contours from the CGPS overlaid. Both HB21 and CTB 1 show the 
typical center-filled X-ray morphology combined with a
shell-like radio morphology that characterizes MM SNRs: in the case of CTB 1,
however, as noted previously the X-ray emission is seen to extend through a gap along the
northeastern portion of the radio shell. 
Figure \ref{ctb1softhard} shows mosaicked images of the
soft and hard X-ray  emission (again corresponding to the energy ranges
E$<$1 keV and E$>$1 keV, respectively); there is a noticeable
difference in the spatial structure between the soft and the hard emission. 
In Figure \ref{ctb1optical} we present an optical H$\alpha$ image of CTB 1 
(courtesy of Robert Fesen) with the contours of the X-ray emission overlaid:
the optical morphology of CTB 1 
is quite similar to the radio morphology with the same shell-like structure
and a prominent gap in the northeast \citep{Fesen97}. Little
H$\alpha$ emission is seen in the interior of CTB 1, nor is any optical
emission seen where the X-ray emission extends through the gap in the
optical and radio shell in the northeast. The observed slight extension of X-ray emission 
in the west through the optical and radio shell is likely either 
residuals due to the broad {\it ASCA} point-spread-function or point sources 
rather than true emission from the SNR.
The {\it ASCA} hard image reveals a point-like source seen in projection against
the diffuse emission of CTB 1: it is located at RA (J2000.0) 00$^h$
01$^m$ 25.$^s$5, Dec (J2000.0) $+$62$^{\circ}$ 29$\arcmin$ 40$\arcsec$ and it lies
very close to the eastern edge of the optical and radio shell.
This source (denoted as 1WGA J0001.4$+$6229) may be
a neutron star possibly associated with CTB 1; we discuss it in 
detail in the next section. In contrast to HB21, a significant 
amount of emission from CTB 1 is seen at energies above 1 keV. 
\par 
Similar to our spectral analysis performed with HB21, we extracted GIS2
and GIS3 spectra from elliptical regions approximately 23$\arcmin$ 
in size from both the southwestern and northeastern portions of the
X-ray emitting plasma. When extracting GIS2 and GIS3 spectra from the
northeastern region, we excluded a region 4 arcminutes in diameter
centered on the position of the discrete X-ray source 1WGA
J00001.4$+$6229 to avoid spectral contamination by this source. 
We also extracted SIS0 and SIS1 spectra from both the
southwestern and northeastern portions of the SNR, again using
square-shaped regions approximately 10 arcminutes on a side.
Unfortunately, the signal-to-noise ratio of the extracted SIS0 and SIS1
spectra for the northeastern region was not sufficient for spectral
analysis and we therefore omitted these spectra from our analysis.
%After background subtraction, the count rates for the GIS2, GIS3, SIS0
%and SIS1 spectra of the southwestern portion of CTB 1 (over the energy
%range 0.6-10.0 keV) were 0.060$\pm$0.001, 0.054$\pm$0.001,
%0.064$\pm$0.001 and 0.043$\pm$0.001 counts per second, respectively. 
We attempted to fit the extracted {\it ASCA} spectra using
the thermal models VAPEC and VNEI along with the model PHABS for the
photoelectric absorption. In contrast to HB21, our fits with either VAPEC 
or VNEI {\it did not} produce a $\chi$$^2_{\nu}$ lower than 1.3 for either 
region and failed to account for the 
hard X-ray emission seen above $\sim$3 keV from both regions.
A two-component model with a soft thermal component with a temperature of 
0.3-0.4 keV (either VAPEC or VNEI) and a second thermal component with a
higher temperature or a power law component was required 
for an acceptable fit to the extracted spectra from both the
southwest and northeast portions of the SNR (see Table \ref{ctb1ascafit}).  
Thus, a hard excess is present in the 
{\it ASCA} spectra of CTB 1: \citet{Lazendic06} also identified a second
thermal component of X-ray emission from this SNR based on analysis of extracted
{\it ASCA} spectra but only for the southwestern region.
%this is the first time that 
%hard X-ray emission has been detected from this SNR. 
This hard X-ray
emission was not detected previously because the X-ray observatories
employed in prior observations of CTB 1 (such as {\it ROSAT}) lacked
the required sensitivity at high energies. 
%As described below, to help 
%constrain the spectral properties of the softer thermal component of
%the X-ray emission as observed by {\it ASCA}, we first analyzed several 
%extracted spectra obtained from the dataset of the archival {\it Chandra} 
%observation of CTB 1. This observation sampled a region corresponding to 
%the interior of the radio and optical shell and corresponds to the {\it
%ASCA} observation of the southwest region of the X-ray emission (see 
%Figure \ref{ctb1asca}). Once
%the properties of the softer component were constrained through the
%analysis of the {\it Chandra} spectra, we could generate informed
%spectral fits of the {\it ASCA} spectra to help constrain the spectral
%properties of the harder component of the X-ray emission. 
\par
\begin{figure}
\epsscale{1.2}
%\plotone{ctb1_chandra_threecolor_mono.ps}
\psfig{figure=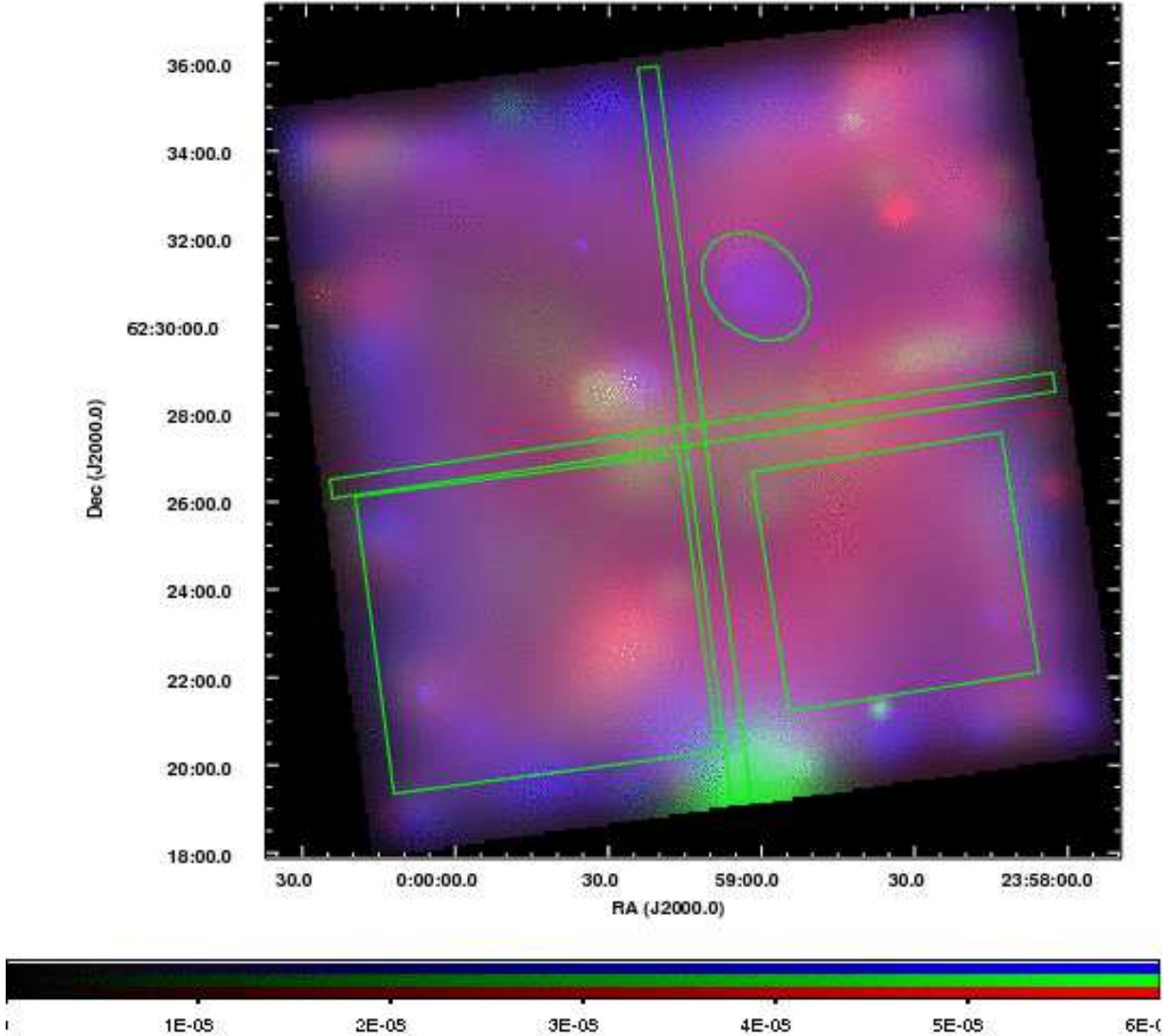,height=15truecm}
\caption{Three-color exposure-corrected and smoothed {\it Chandra}
image of the interior of CTB 1. This image has been made using 
monoenergetic exposure maps. Red, green and blue colors
correspond to soft (0.5-1.0 keV), medium (1.0-2.0 keV) and
hard (2.0-8.0 keV) emission, respectively. The three regions (in green) of spectral
extraction are indicated: clockwise from upper left, these are the
regions denoted as ``hard," (I1 chip), ``soft" (I3 chip) and ``diffuse" (I2 chip) 
excluding point sources 
(see Section \ref{CTB1SubSection}). 
The boundaries of the {\it Chandra} chips are marked with thin boxes (in 
green). Note how clumpy features with different
spectral properties can be distinguished by noticeable differences
in their colors. See Section \ref{CTB1SubSection}.} 
\label{ctb1chandra}
\end{figure}

We examine first the high-spatial resolution {\it Chandra} data before
discussing the hard emission in more detail. 
In Figure \ref{ctb1chandra} we present a three-color {\it Chandra} image 
of the interior X-ray emission surrounded by the well-defined optical and
radio shell of CTB 1. To illustrate the spectral properties of this 
emission, in this Figure we have depicted soft (0.5-1.0 keV), medium 
(1.0-2.0 keV) and hard (2.0-8.0 keV) emission in red, green and blue, 
respectively. Many features with different spectral properties are
visible: most importantly, features with primarily medium and hard
spectra are clearly mixed together within the X-ray plasma. 
At the angular resolution of {\it Chandra}, the hard 
emission (such as the structure $\sim$ 1 arcminute in
size located at approximately RA (J2000.0) 23$^h$ 59$^m$ 01.0$^s$, Dec
(J2000.0) $+$62$^{\circ}$ 30$\arcmin$ 57$\arcsec$) is clearly diffuse and
not point-like. 
\par
To investigate variations in the spectral properties of the X-ray-emitting
plasma of CTB 1 as revealed by {\it Chandra}, we extracted spectra
from three different {\it Chandra} chips. These three extraction regions 
are as follows: the first region (which we refer to as the ``diffuse" 
region) is on the ACIS-I2 chip and 
covers most of the area of this chip. A second region (which we
refer to as the ``soft" region) is located on the ACIS-I3 chip and covers most 
of the area of that chip as well. Lastly we extracted spectra from a third 
region (which we will call the ``hard" region) which corresponds to a region 
of hard emission mentioned above and is located on the ACIS-I1 chip: 
the positions of all three regions are indicated in Figure \ref{ctb1chandra}.
A region of excess medium energy seen toward the middle of the field 
in Figure \ref{ctb1chandra} largely fell into gaps between the {\it Chandra} detector chips:
for this reason, a detailed analysis of its X-ray spectrum could not be conducted.
\par
\begin{figure}
\epsscale{1.0}
%\plotone{ctb1chandraspec.ps}
%\psfig{figure=chandractb1threeregionspec.ps,angle=270,height=11truecm,width=14truecm}
%\psfig{figure=rhochandrathreebest.ps,angle=270,height=11truecm,width=14truecm}
\epsfig{figure=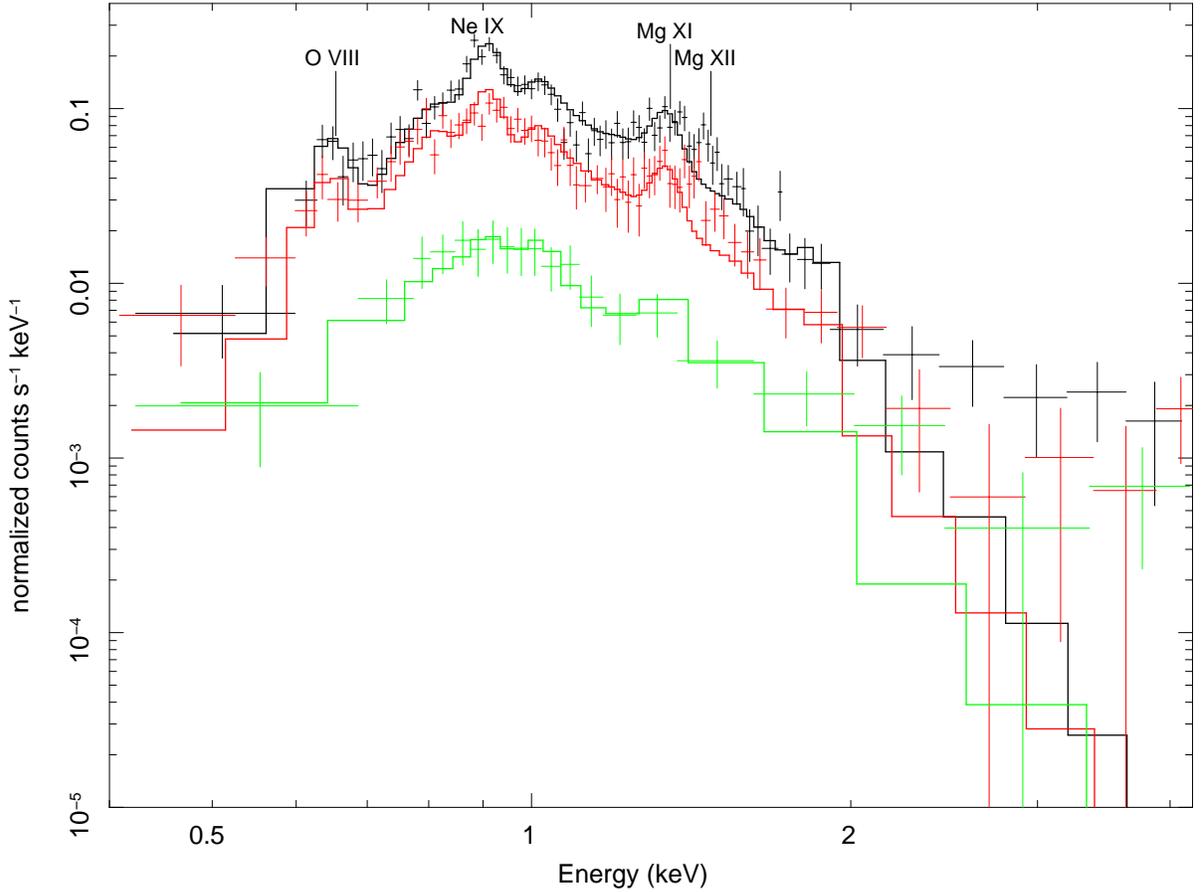,angle=270,width=18truecm}
\caption{{\it Chandra} spectra of the three different regions of
CTB 1. The black, red and green lines correspond to spectra from the ``diffuse," ``soft" 
and ``hard" regions, respectively. The spectra 
have been fit with a PHABS$\times$VAPEC model with the temperature of 0.28, 0.28 and 0.66 keV,
respectively.
%for the ``diffuse" region and the ``soft" region
%have been fit with a PHABS$\times$VAPEC model while the spectrum for the ``hard"
%region has been fit with a PHABS$\times$POWER LAW model. 
The detailed parameters for
these fits are given in Table \ref{ctb1chandrafit}.\label{ctb1chandraspec}}
\end{figure}
\begin{figure}
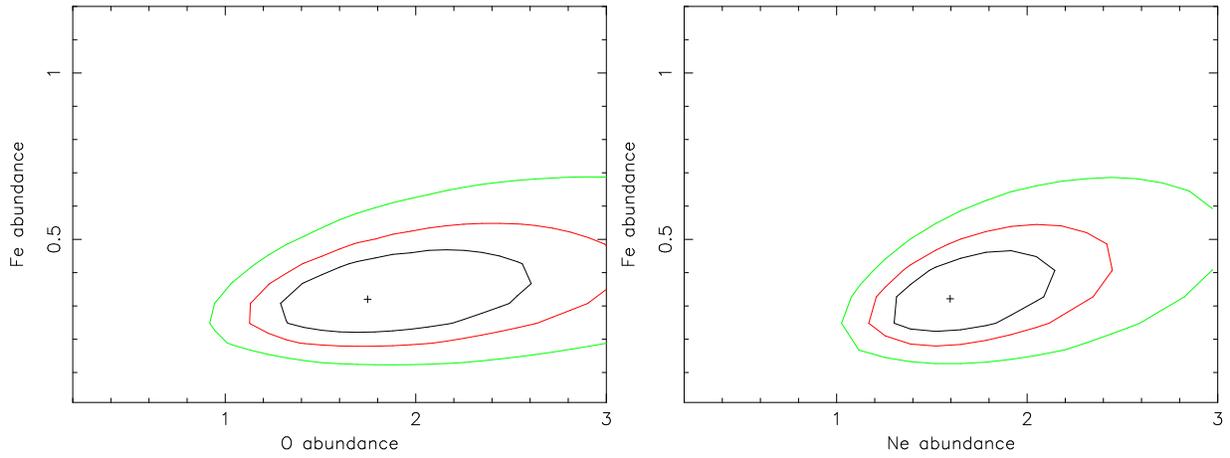

\epsscale{1.0}
\hbox{
\psfig{figure=f12a.ps,angle=270,width=8truecm}
\psfig{figure=f12b.ps,angle=270,width=8truecm}
}
\caption{Confidence contours of O and Fe abundances and Ne and Fe abundances, respectively,
showing enhanced O and Ne abudances and lower Fe abundance relative to solar, for the
PHABS$\times$VAPEC fit to the spectrum extracted for the ``diffuse" region. 
The confidence contours are at the 1$\sigma$, 2$\sigma$ and 3$\sigma$ levels.
See Table \ref{ctb1chandrafit}.
}
\label{ctb1OFeconf}
\end{figure}

 We present the extracted spectra of all three 
regions in Figure \ref{ctb1chandraspec}: spectral variations are present across
the X-ray emitting plasma of CTB 1. For example, the spectrum from the
``diffuse" region shows prominent O Ly 
$\alpha$ (0.65 keV), Ne IX (0.9 keV), and Mg XIII (1.35 keV) lines, together 
with Fe L-shell line emission: in this spectrum the O and Ne lines are stronger
than the Fe L-shell line emission. 
In contrast, the lines are hardly noticeable in the ``hard" region spectrum. 
We derived an acceptable fit to the spectrum of the diffuse region
using a thermal model (VAPEC) with a temperature ${\it kT}$ = 
0.28$\pm$0.03 keV and $N$$_H$ = 0.64$\pm$0.08$\times$10$^{22}$ cm$^{-2}$; this
fit reveals enhanced oxygen and neon abundances and a low iron abundance (see 
Table \ref{ctb1chandrafit}). The column density 
$N$$_H$=0.64$\pm$0.08 $\times$ 10$^{22}$ cm$^{-2}$ is consistent with the optical
extinction of E(B-V)=0.7-1 derived by \citet{Fesen97}, which is equivalent 
to $N$$_H$$\sim$0.5-0.7 $\times$ 10$^{22}$ cm$^{-2}$. 
In Figures \ref{ctb1OFeconf}a and \ref{ctb1OFeconf}b, we present
confidence contour plots for the abundances of oxygen compared to iron and neon
compared to iron (respectively) based on the fit derived from the PHABS$\times$VAPEC
model to the spectrum of the ``diffuse" region. These figures show that the oxygen and
neon abundances are both above solar while the iron abundance is below solar:
such relative abundances are a typical characteristic of oxygen-rich SNRs
\citep{Woosley95}.
\par
Our detection and analysis presented here of probable oxygen-rich ejecta in CTB 1
is the first detailed study presented 
of oxygen-rich ejecta associated with an MM SNR. Based on X-ray spectral analysis, 
\citet{Lazendic06} suggested that another MM SNR -- HB 3 -- may also feature oxygen-rich
ejecta, though those authors could not determine if the plasma associated with that SNR
has significantly enhanced abundances of oxygen, neon and magnesium or marginally
enhanced abundances of magnesium and underabundant iron. 
%no other known MM SNR shows X-ray spectr
% (IS OPTICAL SPECTRA SHOWING O-RICH?)
%Note that Fe L shell models are not perfect due to the uncertainty of
%the current models, but it was sufficient for our spectra which has a
%somewhat low signal-to-noise ratio. 
The {\it Chandra} spectra are not well fit by a
single temperature thermal model (see the poor match to the
Mg XII Ly$\alpha$ line as seen in Figure \ref{ctb1chandrafit}), 
so it is not possible to have oxygen, 
neon and magnesium all residing in the same constant temperature plasma in
ionization equilibrium. Magnesium appears
to be more ionized than neon. This may be caused by higher 
temperatures in magnesium-rich ejecta than in the oxygen- and neon-rich 
ejecta. 
%**OVER-IONIZATION
It is also possible that the high temperature plasma could be
overionized; an overionized plasma has been reported in
the MM SNR IC 443 \citep{Kawasaki02}.
%It may be helpful in the future to model recombining plasmas, but given our 
%limited signal-to-noise of the {\it Chandra} dataset considered here, a 
%new model would help much.
When we fit the spectrum of the 
``soft" region using a VAPEC model combined with
the PHABS model for the interstellar photoelectric absorption,
the iron abundance is higher and the neon abundance is lower when compared with
the ``diffuse" region spectra (see Table 
\ref{ctb1chandrafit}), suggesting spatial variations in the chemical composition
within the X-ray emitting gas of CTB 1. 
%This fit is consistent with the spectra we observed 
%where emission from the Fe line at $\sim$0.8 keV is stronger than the emission
%seen in the ``diffuse"
%region spectra as shown in Fig. \ref{ctb1chandra}. 
%Statistically-acceptable fits (with reduced  $\chi$$^2$ values
%corresponding to unity) are derived in both cases: we varied the
%abundances of oxygen, neon and iron while leaving the abundances of
%other elements frozen at solar values. The derived column densities,
%temparatures and elemental abundances derived for these two regions are
%broadly consistent with each other: some differences between the fits
%may be accounted for as simply differences in the plasma conditions
%between these two sampled regions of CTB 1. Like the  situation with
%our analysis of the HB21 spectra, 
\par
Lastly, we describe the 
spectral analysis of the third (``hard") region which features a harder 
spectrum when compared to the other two spectra (see Figure
\ref{ctb1chandraspec}).
% which shows indeed a higher temperature than the other two regions
%as shown in the confidence contour comparison in Figure \ref{ctb1confidencecontour}. 
%In summary, we can not determine if this hard emission has non-thermal
%nature or thermal nature, but the fits quantified a harder nature of this localized emission
%as we saw in the contrast in colors (Figure \ref{ctb1chandra}). 
%Since there is some residual of line 
% and there is some residual of lines. This suggests that there is
%contribution from diffuse thermal emission. 
%We used multiple models
%to fit the spectrum of this source -- such as a simple power law and
%VAPEC -- as well as combinations of
%models, such as a power law combined with APEC (that is, a thermal
%emission plasma model with abundances frozen to solar) or a combination 
%of two VAPEC models. 
Such localized hard emission has been reported in several 
other MM SNRs, either as non-thermal diffuse knots seen in  
{\it XMM-Newton} observations of IC 443 \citep{Bocchino03} or as a
pulsar wind nebula (also in IC 443 -- \citet{Olbert01}). 
Because of the lack of lines
in the spectrum, we first fit the spectra with a power law, which resulted in
an acceptable fit, but with an unrealistically high photon index of 
$\Gamma$$\sim$6.4$^{+5.6}_{-2.9}$ (see Table \ref{ctb1chandrafit}). 
In comparison, a VAPEC thermal model yielded a comparable-quality
fit with a temperature of $\it{kT}$ = 
0.66$^{+0.27}_{-0.41}$ keV: for this fit, we froze the abundance of oxygen
to 1.7 (to be consistent with the fits derived to the spectra of the 
``diffuse" and ``soft" regions) or 1 while fitting for the abundances of neon
and iron. 
We also fit the spectrum of this region with a 
combination of an APEC thermal component and a non-thermal power-law 
component; to reduce the number of free parameters we 
fixed the temperature of the thermal component to ${\it kT}$ = 0.28 keV,
equal to the temperature derived from fits to the other two regions. The
results are 
given in Table \ref{ctb1chandrafit}: unfortunately, we do not have enough 
counts in the spectrum of the ``hard" region
to distinguish between non-thermal and thermal interpretations
of the spectrum of this source. However, the temperature derived from
the thermal fit ($kT$ = 0.66 keV) is significantly
higher than the temperature of the ``soft" and ``diffuse" regions. Also,
thermal models systematically underpredict {\it Chandra} spectra at high 
energies in all the regions: we interpret this result as evidence 
that the hard excess observed toward CTB 1 has a diffuse origin.
\par
\begin{figure}
\epsscale{1.0}
\psfig{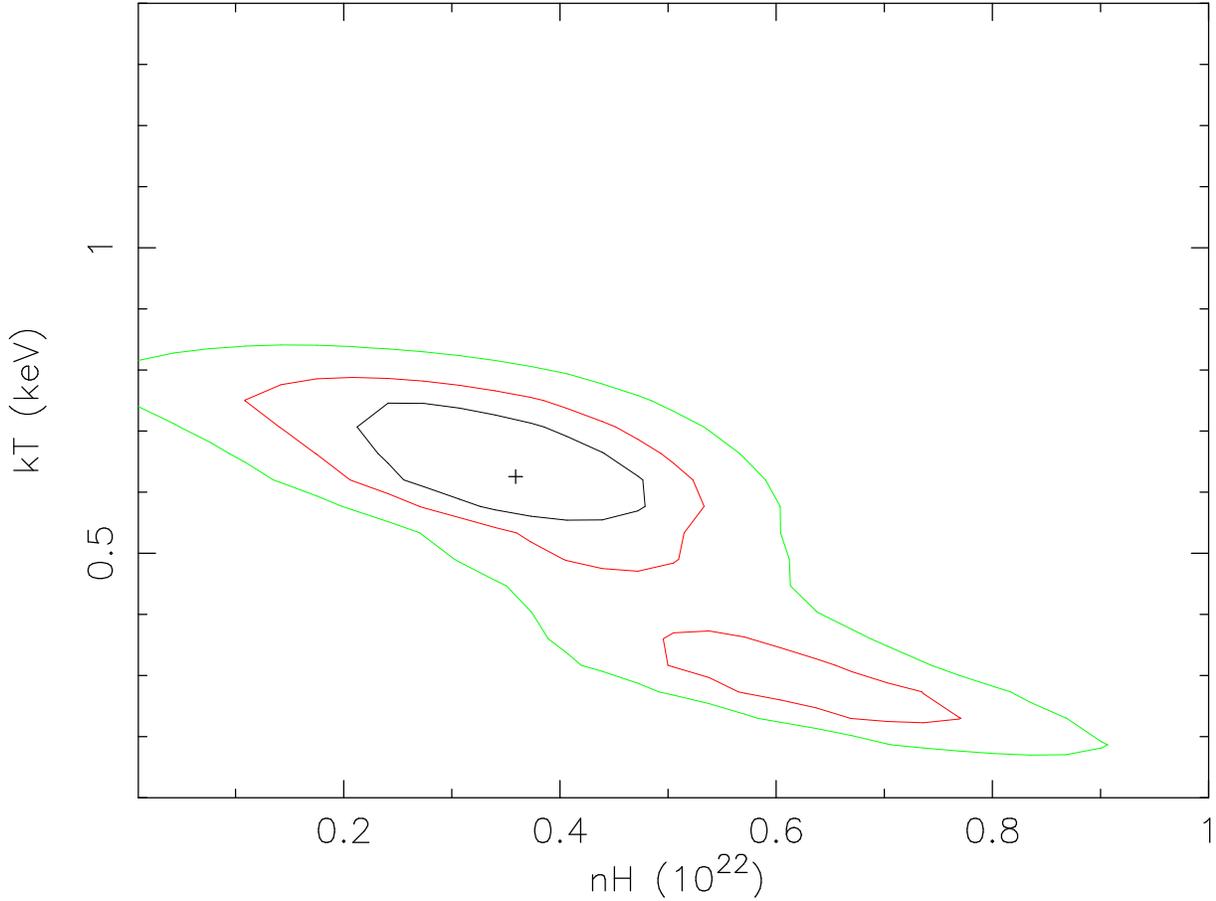}
\caption{
Confidence contours for
the line-of-sight column density $N$$_H$ and the temperature $kT$
for the PHABS$\times$VAPEC spectral fit (with fixed solar abundances) to the spectrum of
the ``hard" region
(see Table \ref{ctb1chandrafit}). Notice the bimodality for components with
temperatures of $kT$ $\sim$ 0.28 keV and $kT$ $\sim$ 0.66 keV: this may
indicate a presence of an additional thermal component besides the
thermal component identified in the fit of the spectrum of the ``diffuse"
region. The confidence contours are at the 1$\sigma$, 2$\sigma$ and 3$\sigma$ levels.
See Section \ref{CTB1SubSection}.
%Confidence contours  of the light of sight absorption (N$_H$) and temperatures 
%(kT) for the spectral fit of the ``hard" region,
%shows bimodality with 0.28 and 0.6 keV temperature
%components. It may indicate an additional component to that of ``diffuse" region.
%[UPDATE THE PLOT IF POSSIBLE. 
%SHOW BOTH CONTOURS OF HARD AND DIFFUSE REGIONS. I HAVE AN
% IDL PROGRAM TO COMBINE IF YOU GIVE QDP,PCO FILES
%FOR DIFFUSE REGION. (IT IS OK TO SKIP THOUGH).]}
}
\label{harddiffuseconf}
\end{figure}

In Figure \ref{harddiffuseconf}, we present a plot of confidence contours for 
$N$$_H$ and $kT$ that correspond to the fit to the spectrum of the ``hard" region as
fit with the PHABS$\times$VAPEC model with fixed solar abundances. The spectrum
of this region demonstrates a bimodality with temperatures of $kT$ $\sim$ 0.28 and 
$kT$ $\sim$ 0.66 keV, suggesting the presence of an additional component in addition 
to the thermal component associated with the ``diffuse" region.
We will discuss the nature of this hard component further below and in 
Section \ref{CTB1HardSection}. We also note differences between our results from analyzing
{\it Chandra} ACIS-I spectra and the results presented by \citet{Lazendic06}: those
authors jointly fit four spectra taken from each ACIS-I chip (they did not attempt any spectral
analysis of small regions  like the ``hard" region") and presented an acceptable fit
obtained using two VNEI components: one with a temperature $kT$ = 0.20$^{+0.04}_{-0.01}$
keV and solar abundances and the other with a temperature $kT$=0.86$^{+0.03}_{-0.06}$
and an elevated magnesium abundance (Mg=3.1$^{+1.0}_{-0.4}$). Those authors
did comment on the presence of lines associated with oxygen as well as the neon and
iron line blend in the extracted ACIS spectra but they did not present an analysis of the
abundance of those elements.
\par
\begin{figure}
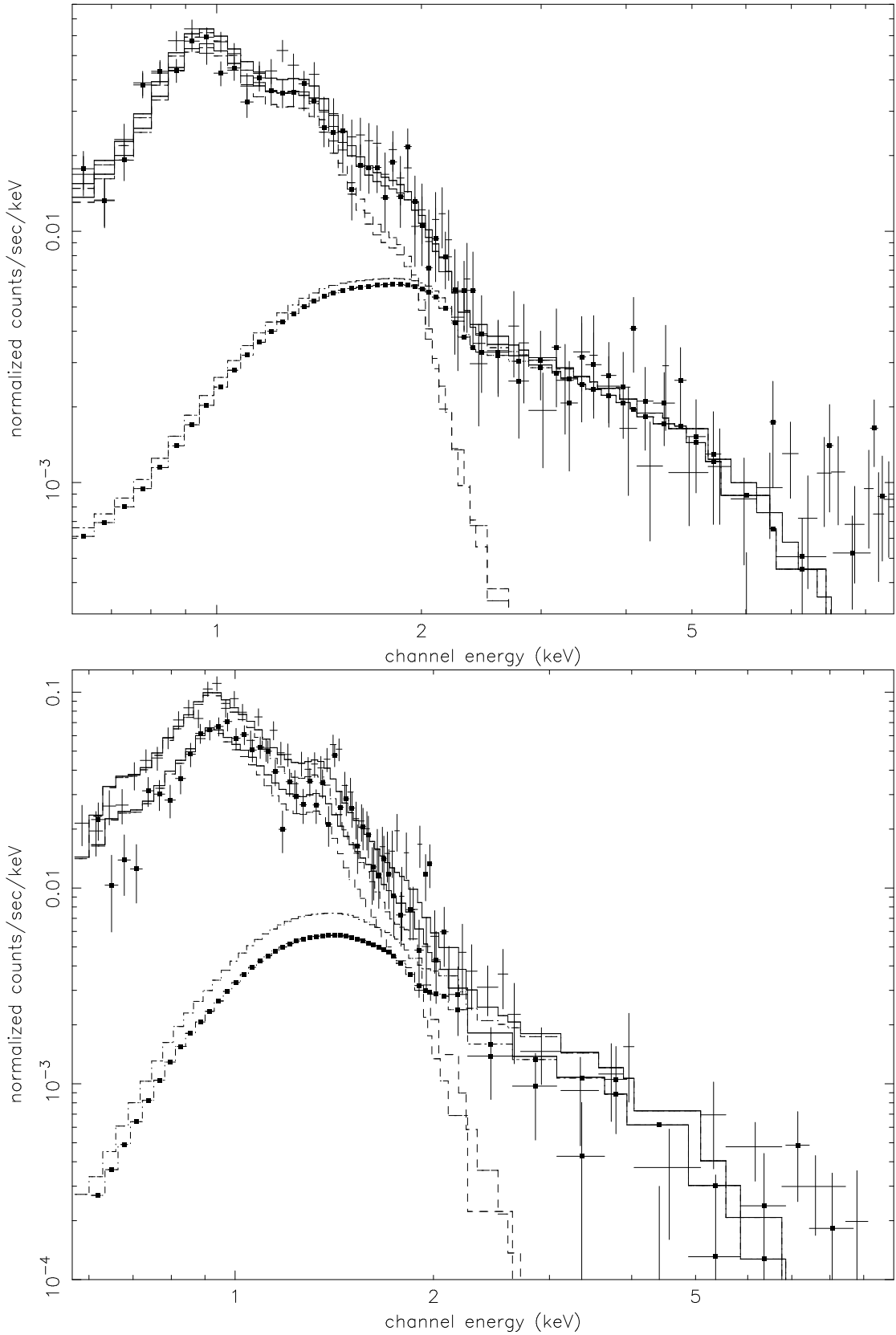

\epsscale{0.85}
\vbox{
\psfig{figure=f14a.ps,width=11truecm,angle=270}
\psfig{figure=f14b.ps,width=11truecm,angle=270}
}
\caption{{\it ASCA} GIS (top) and SIS (bottom) spectra of the southwestern
region of CTB 1: the spectra have been fit using the 
PHABS$\times$(VAPEC+Power Law) model (see Table \ref{ctb1ascafit}).}
\label{ctb1swspectra}
\end{figure}

After establishing the spectral properties of the soft X-ray emitting plasma
from {\it Chandra} spectra, we fit the {\it ASCA} GIS2/3 and SIS0/1 spectra
of the southwestern portion of CTB 1. We froze the oxygen abundance to
1.7 (or 1) for all spectral fits while allowing the neon and iron abundances
to vary. Because models with a single thermal component were not sufficient to
describe the X-ray spectra, we used a combination of the thermal models
VAPEC and VNEI along with a power law model to jointly fit the {\it ASCA} 
spectra: the results of these spectral analyses are given in Table 
\ref{ctb1ascafit}. From these 
fits, we estimate a column density $N$$_H$ $\sim$ 0.5-0.6 $\times$ 
10$^{22}$ cm$^{-2}$ and a temperature $\it{kT_{\rm{soft}}}$ 
$\sim$ 0.2-0.3 keV for the soft emission. The soft 
thermal component has most likely attained 
thermal equilibrium because fits to this soft emission using the VNEI
model resulted in a long ionization timescale ($\tau$ $\sim$
1 ($>$0.2) $\times$ 10$^{11}$ cm$^{-3}$ s). The hydrogen column density $N$$_H$ 
and the temperature of the soft component $\it{kT_{\rm{soft}}}$ are in
agreement with the analysis of {\it ROSAT} spectra \citep{Rho95, Craig97}.
The inclusion of a second component was necessary for obtaining
statistically-acceptable fits to the {\it ASCA} spectra of the southwestern
region (with values for the $\chi$$^2_{\nu}$ $\sim$1.1): the addition
of a power law with a photon index $\Gamma$ $\sim$ 2-3 or a second
thermal component with a temperature ${\it kT}$ $\sim$ 
3 keV yields fits with a comparable quality. Unfortunately due
to a low number of counts in the spectra at the higher X-ray energies,
we cannot distinguish between different models for the high-energy emission.
In Figure \ref{ctb1swspectra}
we present the GIS and SIS X-ray spectra of the southwest 
region, fit with the combination of a thermal (VAPEC) and non-thermal
(power law) model. Our results differ from \citet{Lazendic06} who fit the
extracted ASCA spectra from this region with two thermal components in CIE: the softer
temperature component featured a temperature $kT$ = 0.19$^{+0.09}_{-0.03}$ keV and
a magnesium abundance fixed at solar while the harder temperature component featured
a temperature $kT$=0.82$^{+0.09}_{-0.06}$ keV and an elevated magnesium abundance
(Mg=2.7$^{+0.9}_{-0.5}$).

\par

\begin{figure}
\epsscale{0.85}
%\vbox{
\psfig{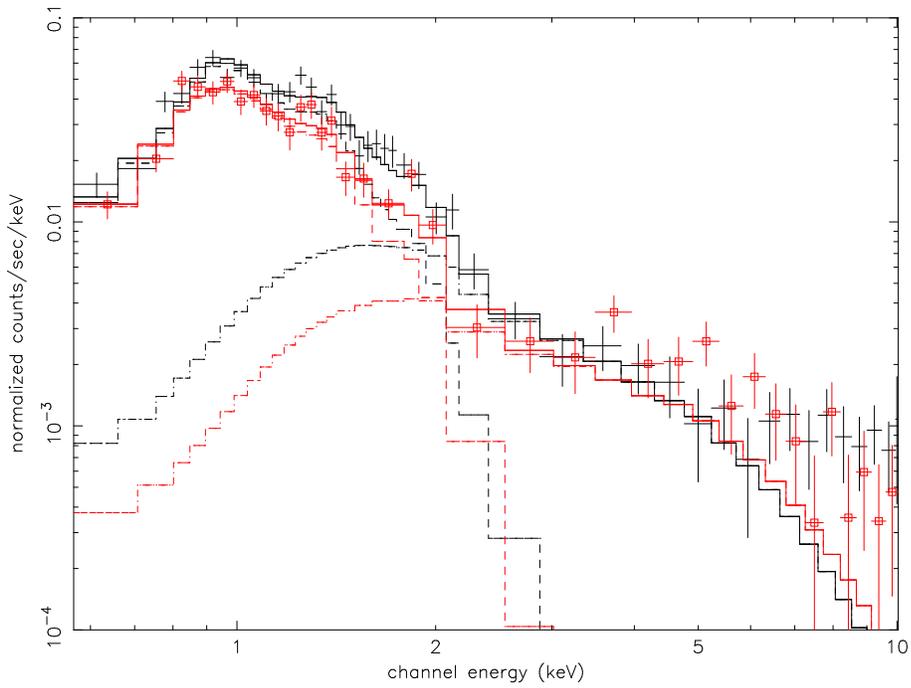}
%\psfig{figure=ctb1tworegionspec.ps,angle=270,width=12truecm}
%\psfig{figure=ctb1tworegionspecldata.ps,angle=270,width=12truecm}
%}
\caption{{\it ASCA} GIS2 spectra of the southwest (in black with crosses) and northeast 
(in red with squares) regions of CTB 1. Both of these spectra have been fit with
a PHABS$\times$(VAPEC+POWER LAW) model. The parameters of the fits to
these spectra are listed in Table
\ref{ctb1ascafit}.\label{ascactb1gisspectra}}
%THIS IS NOT THE BEST MODEL YET.
% (ONLY ONE FIGURE WILL BE INCLUDED IN THE PAPER}\label{fig11}
\end{figure}

Finally, we examine the {\it ASCA} spectra of the northeast region of
CTB 1 which corresponds to the known ``break-out" site seen in optical and radio images
of this SNR.. Like the southwest region, a second component
is needed (in addition to a soft thermal component) to derive a statistically
acceptable fit. A fit to
these spectra using a power-law component for the hard emission is
presented in Table \ref{ctb1ascafit} and the GIS spectra are presented in
Figure \ref{ascactb1gisspectra}. For the thermal component  we have first
assumed the abundances of oxygen, neon and iron to be 1.7, 1.6 and 0.4, respectively,
equal to the abundances in the ``diffuse" region. 
Secondly, we assumed solar abundances for the hard component. Although the 
GIS spectra of the northeastern region feature a stronger Fe L-shell line 
complex when compared with the spectra of the southwestern region, 
our spectral fits could not confirm  
non-solar abundances and the abundances are consistent with solar values.  
\par
The photon index derived from fits to the northeast is flatter than the 
photon index derived from fits for the southwest region: $\Gamma =
1.4$ compared to $\Gamma = 2-3$, respectively. However, the photon statistics
at higher energies is poor, making it difficult to determine
the true nature of this hard emission. Although the northeast region does 
correspond to the prominent breakout
feature, we do not find evidence for any major differences in the X-ray
properties between the northeast and southwestern regions of CTB 1.
The fact that no significant variations are seen on large scales in the X-ray 
properties of CTB 1, coupled with the variations seen on small scales as
revealed by the {\it Chandra} observation, indicate that the X-ray emission
from this SNR is complex. A parallel may be drawn with the results from
{\it Chandra} observations of 3C 391, another MM SNR: in the case of that
source, local spectral differences appeared to be stronger than global ones
\citep{Chen04}. Here again we find that our results differ from
those presented by \citet{Lazendic06}: for this region, the authors derived an acceptable fit 
using a single thermal component ($kT$ = 0.18$^{+0.00}_{-0.01}$ keV) in CIE with solar
abundances. 
%However, the lack of variations may be due to the limited number of counts
%obtained from the northeastern region: a complementary deep {\it Chandra}
%observation of this region may help address this shortcoming.  
%In Table \ref{ctb1ascafit} indeel Higher Fe abundance
%2) possibly somewhat higher temperature
%Si free
%We will discuss the
%nature of the hard emission from CTB 1 in more
%detail in Section \ref{CTB1SubSection}: here, we compare our results
%for our derived column density $N$$_H$ and characteristic temperature
%${\it kT_{\rm{soft}}}$ of the soft component to prior spectral analyses of the
%X-ray emission from this SNR
%: namely, 
%\citet{Craig97} derived comparable numbers for the column density ($N$$_H$ 
%$\sim$ 0.8 $\times$ 10$^{22}$ cm$^{-2}$) and characteristic temperature 
%($\it{kT}$ $\sim$ 0.2 keV) for emission from a plasma with solar abundances. 
%In addition, \cite{Rho95} also derived a statistically acceptable fit using 
%a thermal model with a variable abundance: the derived parameters for that fit
%were $N$$_H$ = 0.49$^{+0.13}_{-0.40}$$\times$10$^{22}$ cm$^{-2}$ and
%${\it kT}$ = 0.2$^{+0.35}_{-0.04}$ keV. We speculate that the presence of
%hard X-ray emission from CTB 1 may at least partially account for discrepancies
%in the fits derived by \citet{Rho95} and \citet{Craig97}.
\section{1WGA J0001.4+6229 -- An X-ray Pulsar Associated with
CTB 1?\label{WGASection}}

%***************************************************************************
%COMPARISON TO CRAB IS NOT REALLY RELEVANT AS THERE IS NO SIGN OF A PULSAR
%WIND NEBULA. DESCRIPTION OF TIMING ANALYSIS IS QUITE OPAQUE FOR A 
%NON-EXPERT --
%EITHER PROVIDE MORE DETAILS OR JUST STATE THAT THE DETECTION IS NOT 
%STATISTICALLY SIGNIFICANT. I DO NOT FULLY UNDERSTAND THE SECOND METHOD FOR 
%ESTIMATING THE NEUTRON STAR TRANSVERSE VELOCITY -- EITHER REFINE EXPLANATIONS
%OR REMOVE MENTION OF THIS METHOD. THE FORMULA FOR THE TRANSVERSE VELOCITY OF
%THE X-RAY SOURCE IS SOMEWHAT MISLEADING AS THE ESTIMATED SNR AGE DEPENDS ALSO 
%ON THE ASSUMED DISTANCE -- DEPENDENCE OF TRANSVERSE VELOCITY ON DISTANCE IS 
%THEN REDUCED. FOR AN ASYMMETRIC REMNANT WITH A BREAKOUT ASSUMPTION OF 
%SPHERICAL SYMMETRY MOST LIKELY IS NOT VALID, AND DERIVED SNR AGES MIGHT BE
%JUST LOWER LIMITS.
%***************************************************************************

The {\it ASCA} hard energy image ($E$ $>$ 1 keV) revealed a hard source in the
northeastern region of CTB 1 which is located just inside the eastern
shell of the SNR: the position of this source is RA (J2000.0) 00$^h$
01$^m$ 25.$^s$5, Dec (J2000.0) $+$62$^{\circ}$ 29$\arcmin$ 40$\arcsec$
with a positional uncertainty of 13$\arcsec$. This is the discrete X-ray 
source 1 WGA J0001.4$+$6229
in the Catalog of {\it ROSAT} PSPC WGA Sources \citep{White94, White97,
Angelini00}\footnote{Also see http://wgacat.gsfc.nasa.gov.}. This X-ray
source may possibly be a neutron star associated with CTB 1.  We
therefore conducted spectral and timing analysis of the X-ray emission
from this source using the {\it ASCA} GIS2 and GIS3 datasets.
\par
\begin{figure}[h]
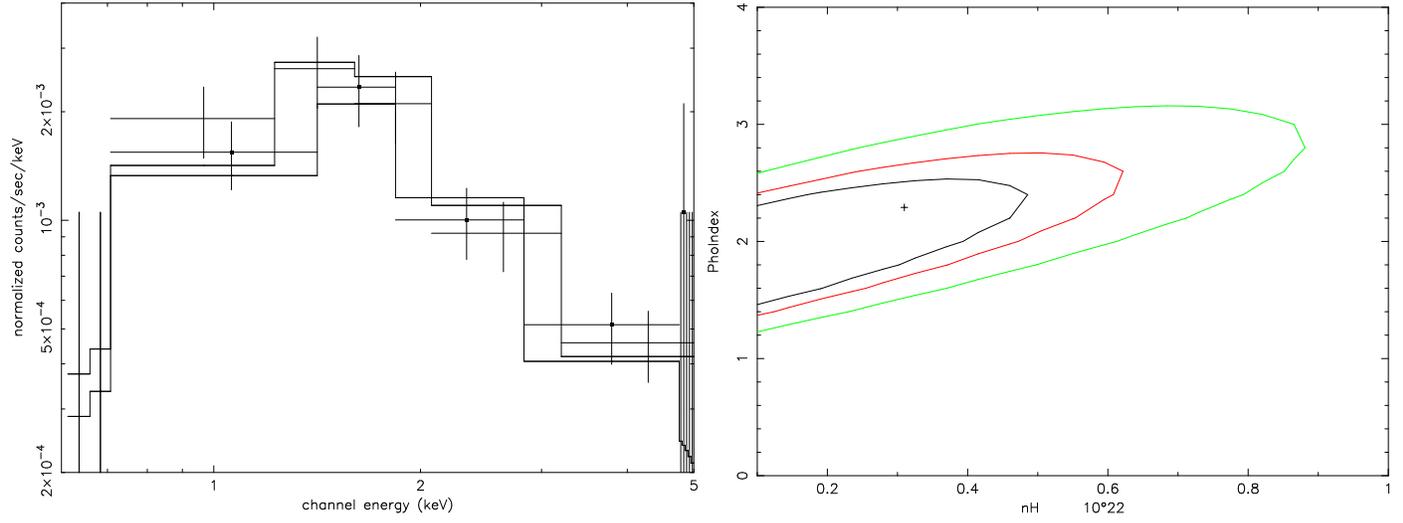

\epsscale{1.0}
\hbox{
\psfig{figure=f16a.ps,width=6.8truecm,angle=270}
\psfig{figure=f16b.ps,width=6.8truecm,angle=270}
}
\caption{(left) The GIS2/GIS3 spectra of the hard source 1WGA J0001.4$+$6229 
as fit with an absorbed power-law model. (right) Plot of confidence contours for Photon Index 
versus $N_H$ for this fit: a low column density is implied, indicating that the source is Galactic
rather than extragalactic. The confidence contours are at the 1$\sigma$, 
2$\sigma$ and 3$\sigma$ levels. See Section \ref{WGASection}.\label{ctb1conf}}
\end{figure}

The procedure for extracting GIS2 and GIS3 spectra of 1WGA J0001.4$+$6229
and performing a spectral analysis was the same as for
extracting GIS2 and GIS3 spectra of the diffuse emission from HB21 and
CTB 1: a circular region
four arcminutes in diameter centered on the source position was used to
extract spectra. The total
number of counts and the corresponding count rate (over the energy
range from 0.6 keV to 10 keV) for our GIS2 and GIS3 observations were
115 and 103 counts, and 5.71$\pm$0.53$\times$10$^{-3}$  and
5.11$\pm$0.51$\times$10$^{-3}$  counts per second, respectively.
We derived a
statistically-acceptable joint fit to the spectra using a simple power
law model combined with the same PHABS model mentioned previously for
the photoelectric absorption along the line of sight. The parameters of
this fit  were a column density of $N$$_H$=0.3 ($<$0.65$)\times
10^{22}$ cm$^{-2}$ and a photon index $\Gamma$=2.2$^{+0.5}_{-1.2}$:
in Figure \ref{ctb1conf} we present the extracted GIS2/GIS3 spectra
together with the best-fit model, and a confidence contour plot for
$N$$_H$ and $\Gamma$. 
%The low column density toward this object and
%lack of Fe K$\alpha$ line emission in the spectra suggest that it is 
%Galactic in origin rather
%than extragalactic: 
%The column density is consistent to those drived the SNR diffuse emission (see Table 
%\ref{ctbascafit} and \ref{ctb1chandrafit})
The derived photon index is typical for 
rotation-powered pulsars. The column density is consistent
with the range of column densities derived in our fits (see Tables 
\ref{ctb1chandrafit} and \ref{ctb1ascafit}) to the 
CTB 1 spectra, hinting at a possible association. 
%In addition, the value of the photon index derived
%from the power-law fit to the extracted spectra is consistent
%with rotation-powered pulsars such as seen in Crab Nebula.
If we fit the extracted spectra for this source with a blackbody model,
a  temperature of kT$\sim$1.1 keV is derived (although this fit with 
$\chi$$^2_{\nu}$=0.92 for 48 degrees of freedom (DOF) is
inferior to the power-law fit with $\chi$$^2_{\nu}$=0.79). Because of the low derived
column density, our estimated absorbed and unabsorbed fluxes for this
source are virtually identical: for the GIS2 (GIS3) spectrum, the flux
is 5.4$\times$10$^{-13}$ (6.4$\times$10$^{-13}$) ergs cm$^{-2}$
sec$^{-1}$; at the assumed distance to CTB 1, these
fluxes correspond to luminosities of 6.2$\times$10$^{32}$
(7.4$\times$10$^{32}$) ergs s$^{-1}$, respectively. We also performed a
timing analysis using the GIS2/GIS3 datasets to search for pulsed X-ray
emission from this source and detected  
a period of 47.6154 milliseconds using the Rayleigh test (a maximum signal 
of $Z$$^2$ = 31.4), but the detection is not statistically significant.  
%ot significant evidence for
%any pulsations: {\it using the Rayleigh test we obtained a pulsation
%with a period of 47.6154 milliseconds
%(a maximum signal of $Z$$^2$ = 31.4; see details of definition of 
%Rayleigh test at Leigh, 1983), but
% this detection is marginal (less than 3$\sigma$ detection)
% so that a false detection of this pulsation can not be ruled
%out.}  
\par
We further searched for pulsations from 1WGA J0001.4$+$6229
with the 100-meter Green Bank Telescope (GBT) of the National Radio
Astronomy Observatory (NRAO\footnote{The National Radio Astronomy
Observatory is a facility of the National Science Foundation, operated
under cooperative agreement by Associated Universities, Inc.}) on 2004
December 7. The
target position was observed for 7.6 hours at a center frequency of 825
MHz. The frontend was the GBT Prime Focus 1 receiver to feed the Pulsar
Spigot \citep{Kaplan05} and Berkeley-Caltech Pulsar Machine (BCPM)
backends. The receiver provided 50 MHz of bandwidth in two orthogonal
polarizations that were summed and synthesized into 1024 frequency
channels every 81.92 $\mu$s in the Spigot and 96 channels of 0.25 MHz
width every 144 $\mu$s in the BCPM. The interstellar dispersion toward
CTB1 is unknown, but we can estimate it with the latest model of
Galactic electron density \citep{Cordes02}, which predicts a dispersion
measure (DM) of 105 pc cm$^{-3}$ for a distance of 3.1 kpc or DM = 33
pc cm$^{-3}$ at a distance of 1.6 kpc (corresponding to the two distances
to CTB 1 that have been published in the literature). We therefore take 
a conservative upper limit
of DM = 1000 to CTB 1 (given the narrow channels and relative long
pulsation period the search is not highly DM dependent). The data set
was dedispersed with DMs from 0 to 1000 and searched for periodicities
using standard folding and fast Fourier-Transform (FFT)-based
techniques. Based on these analyses, we find no significant evidence
for pulsations with any period from 1WGA J0001.4$+$6229.
%\par
%We can put a lower limit to the flux density of $S_{\rm min} < 0.030$ mJy
%assuming a pulse duty cycle of 10\% \citep{Dewey85}. This is fainter by a
%factor of $\sim 4$ than the faintest known young pulsar, J0205+6449
%\citep{Camilo02}.
\par
% Considering that not
%all X-ray pulsars have been detected at radio wavelengths, it may be
%still possible that the X-ray pulsations from this source are real. The
%pulsar may be similar to the pulsars seen in the MM SNRs W44 and IC 443
%(REF): if the pulsation were real, the millisecond period would be
%comparable to period of the pulsar associated with W44,  though 1WGA
%J00001.4$+$62229 is located far from the geometric center of CTB 1. 
Assuming that 1WGA J0001.4$+$6229 is in fact a neutron star associated
with CTB 1, a transverse velocity can be estimated. The angular displacement
of 1WGA J0001.4$+$6229 from the center of CTB 1 is 
14$'$ while the radius of the SNR itself is 17$'$. Therefore, we calculate
%a transverse velocity v$_{PSR}$= 840 d$_{3.1kpc}$
%t$_{15}^{-1}$ km s$^{-2}$, where d$_{3.1}$ is the distance to CTB 1 in units 
%of 3.1 kpc and t$_{15}$ is the age of the SNR in units of 15 $\times$ 
%10$^3$ yr. 
a transverse velocity $v$ = 850 d$_{3.1}$
t$_{1.6}^{-1}$ km s$^{-1}$, where d$_{3.1}$ is the distance to CTB 1 in units 
of 3.1 kpc and t$_{1.6}$ is the age of the SNR \citep{Fesen97} in units of 1.6 
$\times$ 10$^4$ yr.
This estimated transverse velocity is high but this may be
an overestimate because of the considerable uncertainties associated with
estimates of the distance and age of CTB 1: if we assume a distance
to the SNR of 1.6 kpc, the tranverse velocity is only 
%433 km s$^{-1}$.
420 km s$^{-1}$. 
In particular, the published age estimates are based on simple one-dimensional
SNR models: the obvious breakout morphology of CTB 1 clearly indicates
that such models are not applicable in this case. For comparison, the
transverse velocities for neutron stars located off center
in their associated SNRs are 375 km s$^{-1}$ in the case of the SNR
W44 \citep{Frail96} and 250$\pm$50 km s$^{-1}$ in the case of the SNR IC 443
\citep{Olbert01}. 
%Another way to estimate the transverse velocity of the pulsar is
%V$_{PSR}$= $\beta$ R$_s$/t where R$_s$ is the shock radius and t is the pulsar 
%age. With V$_s$ of 107 km s$^{-2}$ estimated by \citet{Koo91}, this estimate 
%of the transverse velocity is consistent with the large transverse velocity 
%just quoted.
%Therefore, it is plausible to have
%a pulsar with this high kick velocity. 
It is plausible that 1WGA J0001.4$+$6229 is associated with CTB 1
but deeper high-spatial resolution X-ray observations are needed to
examine its spectrum in more detail and search for possible pulsations.

\section{Plasma Conditions in HB21\label{HB21SubSection}}

We first estimate the density and mass of the X-ray emitting
plasma associated with HB21 based on the emission measures derived from 
our spectral fitting. Our GIS spectral extraction regions extend over
approximately 11$\farcm$5 $\times$ 11$\farcm$5 or 5.7 pc $\times$
5.7 pc (2.7 pc $\times$ 2.7 pc) at the assumed 1.7 (0.8) kpc distance to HB21. Assuming a
cylindrical geometry with the long axis equal to the observed
extent of the X-ray plasma (35$\farcm$8, corresponding to 17.7 (8.3) pc),
the volume of each region is approximately 5.4$\times$10$^{58}$ (5.6$\times$10$^{57}$)
cm$^{3}$. From the mean values of our derived emission measures (which
are approximately the same for all regions and all
models), we calculate an electron density $n$$_e$ $\approx$ 0.06 (0.08) cm$^{-3}$
(where we have assumed $n$$_e$ $\approx$ 1.2$n$$_H$) and a volume filling
fraction of unity based on the smooth appearance and isothermal nature
of the X-ray emitting gas. Based on this value, we estimate the total mass of the X-ray 
emitting plasma within the field of view of the {\it ASCA} observations to be only 
$\approx$$2.6 (1.5) M_{\odot}$.  When we account for the incomplete spatial coverage of 
HB21 by the {\it ASCA} observations, the total
X-ray mass could be higher by a factor of 9, which amounts to a total of 
23.4 (14) $M_\odot$.  The corresponding Si and S masses are 
$1.7 \times 10^{-3} M_{\odot}$ and
$1.3 \times 10^{-3} M_{\odot}$,  respectively. 
%and the preshock density is only n$_0$=n$_e$/4.8=0.016
%cm$^{-3}$. This number is surprisingly low considering that
%HB21 is known to be interacting with molecular clouds \citet{koo98}:
%the low mass and density also imply 
%that the preshock medium is very tenuous and the molecular
%clouds were present in a localized area. 
% but it is significant less than those
%of 3C391 and W44. Unlike HB21 and W28, both  
%3C391 and W44 are completely surrounded by large molecular clouds. 
The presence of a bright radio shell without associated X-ray emission 
combined with the detections of an expanding HI shell and infrared emission from shock-cloud
interaction regions
\citep[][and the references therein]{Koo91,Shinn09,Shinn10} imply that HB21 is in a 
radiative cooling stage. 
The SNR age inferred from the presence of an expanding
($v_{exp}=124$ km s$^{-1}$) HI shell is $t$$_d$ = 4.5$\times$10$^4$ yr
\citep{Koo91}. We can also infer the pressure within HB21 from properties
of the X-ray emitting gas. The total number of particles is  
$n$$_{total}$ = $n$$_e$+$n$$_H$+$n$$_{He}$$\approx$2$n$$_e$ for a plasma
with cosmic abundances: from our estimated values for electron density
and temperature, the corresponding pressure is P/k = 2$n$$_e$$T$ = 
0.9 (1.2)$\times$10$^6$ K cm$^{-3}$, which as about two orders of 
magnitude higher than the typical ISM pressure.  The physical properties of HB21 are 
summarized in Table \ref{physical}.
\par
The X-ray properties of HB21 are similar to those measured for many other 
MM SNRs. First, the presence of an isothermal plasma with a temperature of $kT$ $\sim$
0.2-0.7 keV is consistent with other MM SNRs such as 3C391 \citep{Rho96, Chen04}, W44
\citep{Rho94, Shelton04}, 3C400.2 \citep{Yoshita01}, W51C
\citep{Koo02}, W63 \citep{Mavromatakis04}, and  Kes 79 \citep{Sun04}.
This result supports the interpretation that 
these SNRs are evolved and in the radiative phase as suggested by
the presence of infrared, optical and HI shells for many of
these SNRs. There are no temperature variations and no pronounced enhancements
of chemical abundances in HB21 either, just as in many other MM SNRs like
3C 391. 
These properties of HB21 exemplify the typical X-ray properties characteristic
of MM SNRs as defined by \citet{Rho98}. At a sufficiently old age ($\sim 10^6$
yr) age, a SNR should exhibit a centrally-filled X-ray morphology and
eventually merge with the hot ISM gas \citep{Cui92}: however, MM SNRs
attain this state at a much earlier age ($\sim 10^4$ yr). 
When a distance of 1.7 kpc to HB21 is assumed, the calculated X-ray 
emitting mass of this SNR is comparable to those of other MM SNRs.
%rather small and is much less than expected from
%a radiative SNR with parameters appropriate for HB21. 
For standard radiative SNR models, we expect 
$\sim 100$ M$_\odot$ of X-ray emitting gas at the HB21 age of 
$4 \times 10^4$ yr based on equations given by \citet{Cui92} and models 
presented by \citet{Hellsten95}. Even more X-ray emitting gas is 
expected in conduction models of \citet{Cox99}. This discrepancy between 
the observed and predicted X-ray emitting mass is also present in W28 
\citep{Rho02}, but unlike in W28 (where large temperature gradients 
have been detected), the presence of an isothermal plasma at the center of HB21 
suggests that the electron thermal conduction is important \citep{Chevalier99}.
The conduction model of \citet{Cox99} overpredicts the mass of 
X-ray emitting gas: however, this model assumes a uniform ambient ISM while HB21 is
known to be interacting with with clumpy molecular clouds \citep{Koo01,Shinn09,Shinn10}. 
More elaborate X-ray emission models 
of SNRs in molecular clouds are needed to account for the observed X-ray 
properties of HB21 and similar MM SNRs.

\section{Supernova Ejecta in CTB 1\label{CTB1EjectaSection}}

The enhanced abundances of oxygen and neon and low iron abundances in 
the ``diffuse" region (see Table \ref{ctb1chandrafit}) indicate that CTB 1 is likely an
oxygen-rich SNR.  This SNR was likely produced by a core-collapse SN explosion,
because such explosions produce O- and Ne-rich, and Fe-poor ejecta 
\citep{Nomoto97,Woosley95}. This finding is consistent with the presence of 
a massive star forming environment near CTB 1 \citep{Landecker82}: in addition, the
scenario for the creation of this SNR by a core-collapse SN explosion would be further
supported if the discrete X-ray source discussed earlier -- 1WGA J0001.4$+$6229 --
is shown to be a neutron star associated with this SNR. 
Examples of oxygen-rich SNRs include young sources like Cas A, N132D,
and E0102.2$-$72.3; recently two older SNRs located in the Small Magellanic Cloud 
(SMC) -- SNR B0049-73.6 
\citep{Hendrick05} and B0103-72.6 \citep{Park03} -- were also classified as
oxygen-rich SNRs. Both of these SMC SNRs show the ejecta material
in their interiors surrounded by shells of swept-up ambient material at 
relatively low X-ray emitting temperatures. An X-ray emitting shell might
be present in CTB 1, but its detection may be prevented by substantial interstellar 
absorption in this direction combined with 
an expected low temperature of the shocked ambient gas.
\par
We estimated the X-ray mass and density of CTB 1 from our fits to the 
{\it ASCA} spectra, assuming metal abundances 
derived from the {\it Chandra} spectra of the ``diffuse" and ``soft" 
regions. The spectral fits to the GIS spectra in the 
11$\farcm$5 $\times$ 11$\farcm$5 region 
imply an electron density of $0.16 f^{-1/2}_{soft}$ cm$^{-3}$ for the soft 
thermal component (where $f_{soft}$ is the volume filling factor for this 
component). The corresponding hydrogen density $n_H$ is equal to 
$n_e/1.2 = 0.13 f^{-1/2}_{soft} {\rm cm}^{-3}$; based on this value we 
estimate the total mass of the X-ray emitting plasma to be $\approx 
40 f^{1/2}_{soft} M_{\odot}$. 
From our derived abundances of oxygen, neon and iron based on {\it
Chandra} data, we estimate the oxygen, neon and iron masses to be 
$0.66 f^{1/2}_{soft}$, $0.11 f^{1/2}_{soft}$, and $0.03 f^{1/2}_{soft}$ 
M$_{\odot}$, 
respectively. The ratio of [O/Fe] is 4.3$^{+10.2}_{-2.5}$ and [Ne/Fe] is 
4.0$^{+8.0}_{-2.2}$ for CTB 1.
The expected ratio of [O/Fe] is 0.75 for a Type Ia explosion and greater 
than 4 for a core-collapse explosion. 
These abundances imply that CTB 1 is a remnant of a core-collapse explosion
and are consistent with the predictions for a stellar progenitor with a mass 
of 13 - 15 M$_{\odot}$ \citep{Woosley95, Nomoto97}, but higher mass stellar 
progenitors are not excluded.
\par
 Finally, we estimate the pressures of the
soft and hard components of the X-ray emitting gas using the parameters of
the PHABS$\times$(VAPEC+VAPEC) model: for the soft component we calculate a 
corresponding pressure P/k = 1.1$\times$10$^6$ $f$$_{soft}^{-1/2}$ K
cm$^{-3}$. For the hard component (assuming a thermal origin), we first
need to calculate the corresponding electron density which we can determine
from the electron density of the soft component and the emission
measures (EMs) of the soft and hard components (i.e., $n_{e}$(hard) =
$n_e$(soft) [EM$_{hard}$/EM$_{soft}$]$^{1/2}$). From this relation, we 
obtain $n$$_e$(hard) = 0.029 $f$$_{hard}^{-1/2}$ cm$^{-3}$
(here $f$$_{hard}$ is the volume filling factor for the hard component) 
and therefore a corresponding pressure P/k = 2.0$\times$10$^6$
$f$$_{hard}^{-1/2}$ K cm$^{-3}$. This result implies a factor of
four larger filling factor for the hotter gas than the cooler gas if
these two components are in pressure equilibrium: a higher filling
factor for the hot gas is typical. Assuming $f_{soft}+f_{hard}=1$, pressure
within CTB 1 is $2.8 \times 10^6$ K cm$^{-3}$.     
We summarize these inferred physical properties for CTB 1 in
Table \ref{physical}. 
\par
%************
%I DO NOT AGREE ABOUT A VERY LOW DENSITY ENVIRONMENT IN CTB1. 
%EXTENSIVE OPTICAL EMISSION ENCIRCLING THE REMNANT TELLS US THAT RADIATIVE 
%SHOCKS ARE PRESENT -- THERE MUST BE ENOUGH OF DENSE MATERIAL TO 
%SLOW DOWN 
%SHOCKS. IT IS AN OPEN QUESTION WHETHER OR NOT THE BLAST WAVE ITSELF BECAME 
%RADIATIVE -- IN ORDER TO ANSWER THAT QUESTION ONE WOULD LIKE TO EXAMINE AN 
%X-RAY EMITTING SHELL, BUT SUCH SOFT X-RAY EMITTING SHELL MIGHT BE HIDDEN BY 
%THE INTERSTELLAR ABSORPTION. NOTE THAT FOR VERY LOW AMBIENT DENSITIES 
%X-RAY 
%EMITTING TEMPERATURES IN THE BLAST WAVE SHOULD BE HIGH ENOUGH TO ALLOW %FOR 
%ITS DETECTION -- ONE WOULD NEED TO SET UP UPPER LIMITS BASED ON EXISTING 
%X-RAY 
%DATA IN ORDER TO EXPLORE THIS ISSUE FURTHER. BASED ON THE X-RAY %MORPHOLOGY 
%SEEN IN THE IMAGES, I EXPECT A RATHER ADVANCED DYNAMICAL STATE FOR THIS 
%REMNANT, WHICH FAVORS A RELATIVELY DENSE ENVIRONMENT AND A RADIATIVE %REMNANT. 
%AGAIN, THIS IS JUST MY IMPRESSION, ONE WOULD HAVE TO INVESTIGATE THIS IN MORE 
%DETAIL.BUT I DO NOT THINK THAT MODELING OF THIS REMNANT SHOULD BE OUR GOAL, 
%PERHAPS IT MIGHT BE ENOUGH JUST TO STATE THE FOLLOWING:
%******************************

CTB 1 therefore belongs to a growing number of known evolved SNRs which feature
an enhanced metal abundance in their interiors. An example of another MM SNR
which features such enhanced abundances is W44 \citep{Shelton04}: other 
similar sources are identified by \citet{Lazendic06} 
(including HB21, which was analyzed both in their study and in the
study presented here.)
In addition, two other Galactic SNRs -- the Cygnus Loop \citep{Miyata98}
and G347.7$+$0.2 \citep{Lazendic05} -- feature enhanced abundances of metals as well.
W49B  \citep{Hwang00} shows highly enhanced abundances but its age is estimated to be
2000 years \citep{Hwang00} and \cite{Rho98} describe the source as an atypical MM SNR. 
Because MM SNRs like CTB 1 are commonly believed to be evolved sources -- 
age estimates of CTB 1 range from 9000 yr \citep{Craig97} to 4.4$\times$10$^4$ yr 
\citep{Koo91} -- their X-ray spectra are dominated by swept-up material \citep{Rho98}. Therefore,
 the detection of X-ray-emitting material associated with these sources with enhanced metal 
abundances is unexpected. The detection of O-rich ejecta associated with CTB 1 is
particularly noteworthy: CTB 1 may belong to a previously unrecognized class of MM SNRs 
whose X-ray emission is dominated by O-rich ejecta located within
their interiors. As noted previously, another possible member of this particular class of 
MM SNRs with O-rich ejecta may be
HB3 \citep{Lazendic06}. We note that O-rich ejecta has been previously detected in the evolved 
(1.4 $\times$10$^4$ yr old) SMC SNR B0049-73.6 by \citet{Hendrick05} and it is likely 
that centrally-located ejecta will be found in a number of relatively old
Galactic SNRs.  
\par
The X-ray emitting plasma associated with CTB 1 clearly extends through 
the gap in the crescent-shaped radio 
shell: \citet{Hailey94} and \citet{Rho95} first noticed this remarkable 
extension of X-ray emission based on {\it ROSAT} PSPC observations.
Two scenarios have been proposed to explain the morphology of the
observed X-ray emission: \citet{Craig97} has suggested that the
ambient ISM toward the northeastern portion of the
SNR was cleared by a supernova event which took place prior
to the birth of CTB 1, and thus
a breakout occurred as the SNR expanded into this region of
a dramatically lower density.  
A competing theory for the morphology has been proposed by
\citet{YarUyaniker04}, who suggested that the X-ray emission from CTB 1
lies in the interior of a bubble seen in the 21 cm H line, presumably blown by 
winds of the CTB 1 stellar
progenitor. Our X-ray images show that the diffuse X-ray
emission in the northeast clearly extends through the relatively narrow 
break in the optical and radio shell, with the breakout directed into the 
interior of the bubble seen in the neutral hydrogen line. Such a morphology 
favors a scenario where a supernova explosion occurred within the HI shell,
followed by subsequent breakout into 
the bubble and not an explosion within the bubble itself.
Our {\it ASCA} spectra show little difference 
between the X-ray properties of the southwest and northeast regions: 
hints of variations in temperature and abundances exist but
better X-ray data are needed to determine whether they are real and not just 
statistical fluctuations.

\section{The Nature of the Hard X-ray Emission from 
CTB 1\label{CTB1HardSection}}

The {\it ASCA} spectra of the southwestern portion of CTB 1 revealed 
the presence of a hard component in addition to the soft ($kT$ $\sim$ 0.28 keV) 
component. This hard  
component may be modeled as a second thermal component with a temperature of 
$\it{kT}$ $\sim$ 3 keV or as a power-law continuum with a photon index 
$\Gamma$ $\sim$ 2-3. 
Hard X-ray emission was also detected by {\it ASCA} in the northeast
region of CTB 1: a power-law component with a photon index $\Gamma$
$\sim$ 1.4 (a somewhat lower value compared to the southwest region) combined
with a soft thermal component (again with a temperature $\it{kT}$ $\sim$ 0.28 keV)
yields a statisically acceptable fit. The {\it Chandra} observation of the 
southwest region of CTB 1
revealed regions of harder emission patches on the scale of an
arcminute in size: the spectrum of one of these regions can be modeled by 
either a single thermal component with an elevated temperature ($\it{kT}$ 
$\sim$ 0.66 keV) or as the combination of a soft thermal component 
($\it{kT}$ $\sim$ 0.28 keV) and a power law component with a photon index 
$\Gamma$ $\sim$ 2.0 (see Tables \ref{ctb1chandrafit} and
\ref{ctb1ascafit}).  The ``hard" region observed by {\it Chandra} 
is only a few arcminutes from the center of CTB 1 and well inside the 
radio-emitting shell of the SNR: it is diffuse in nature although the number 
of counts detected from the source is limited. It is possible that the
hard X-ray emission detected by {\it ASCA} from CTB 1 may be composed of 
localized hard regions such as this one: unfortunately we do not have
enough counts in this ``hard" region to distinguish
between thermal and non-thermal origins. 
We note that two other MM SNRs, W28 and IC 443, contain high-temperature
thermal plasmas in their interiors \citep{Rho02, Kawasaki02}.
\par
Several possible explanations may
be considered for the origin of hard X-ray emission
from MM SNRs: first, the hard emission may be caused by 
temperature variations within the SNR. In the case of CTB 1, this scenario
is supported by a good fit to the spectrum of the ``hard" region 
with a thermal component with a much higher than average temperature. 
Supernova ejecta may be inhomogeneous, in which case a multi-temperature
plasma with spatially-varying abundances is expected. 
%This result is
%in contrast to the X-ray spectral analysis of other MM SNRs like W28 and
%3C 391. 
Alternatively, the ``hard" regions may be caused by localized nonthermal
emission: such emission has already been detected in IC 443 \citep{Bocchino03} 
and $\gamma$ Cygni \citep{Uchiyama02}.
%In the second scenario, the hard component is associated with ejecta 
%that has an intrinsially higher temperature: this scenario would help 
%explain the localization of regions of hard emission approximately one
%arcminute in size (corresponding to one parsec at the assumed distance to 
%CTB 1) observed to mixed with the rest of the X-ray emitting plasma 
%associated with the SNR. Observations of several other SNRs such as 3C
%397 \citep{SafiHarb00} and Sgr A \citep{Maeda02} have revealed the presence
%of Fe-rich knots associated with these SNRs; perhaps the observed 
%``hard" region of CTB 1 is a Fe-rich knot as well. In the final scenario
%the ``hard" regions actually correspond to localized regions of non-thermal
%hard X-ray emission: such regions have already been detected in the X-ray
%emitting plasma associated with the SNRs IC 443 \citep{Bocchino03} and
%$\gamma$ Cygni \citep{Uchiyama02}.
Additional observations are needed to understand the true nature of the hard 
X-ray emission from CTB 1. 

\section{Summary}

1. We presented {\it ASCA} observations of the MM SNR HB21. Our
{\it ASCA} images of this SNR are similar to {\it ROSAT} images and 
reveal a diffuse centrally filled X-ray emission located 
within a radio shell. From X-ray spectra, we measure a column density
toward this source and a temperature for the X-ray emitting plasma of 
$N$$_H$ $\sim$ 0.3$\times$10$^{22}$ cm$^{-2}$ and
$\it{kT}$ $\sim$ 0.7 keV, respectively: no significant spatial
differences in temperature are found. 
Silicon and sulfur abundances are          
slightly enhanced relative to solar, particularly for the northwestern region, and
no hard component to the X-ray emission was detected.
The properties of HB21 are similar to those seen in several other MM SNRs,
such as the presence of isothermal plasma. This result supports the
interpretation that MM SNRs are evolved sources currently in the radiative 
phase of evolution: the X-ray properties of HB21 exemplify the primary
characteristics of MM SNRs as defined by \citet{Rho98}. 
\par
2. We presented {\it ASCA} and {\it Chandra} observations of the MM SNR CTB 1. 
{\it ASCA} observations reveal center-filled X-ray emission located  
within the radio shell: the X-ray emission extends outside 
the circular shell through the breakout
gap in the northeast. While the global X-ray and radio morphology is
similar to HB21, the X-ray spectra of CTB 1 and HB21 are very different. The 
X-ray spectrum of CTB 1 shows several prominent lines such as O Ly$\alpha$ 
(0.65 keV) and Ne IX (0.9 keV). We find that CTB 1 is likely an oxygen-rich SNR 
with enhanced abundances of oxygen and neon: this is surprising for 
an evolved SNR such as CTB 1. 
The derived abundances are consistent with an explosion of a stellar 
progenitor with a mass of 13 - 15 M$_{\odot}$ and possibly even higher. 
\par
3.  The {\it ASCA} spectra of the southwest region of CTB 1 cannot be fit 
with a single thermal
component and instead require the presence of an additional component
to account for an excess emission seen at higher energies. Based on
{\it ASCA} and {\it Chandra} spectra of CTB 1, we derive a column
density $N$$_H$ $\sim$ 0.6$\times$ 10$^{22}$ cm$^{-2}$ and the soft
component temperature  $\it{kT}$$_{soft}$$\sim$ 0.28 keV; the hard
emission may be  modeled either by a thermal component with a
temperature $\it{kT}$$_{hard}$ $\sim$ 3 keV or by a power law component
with a photon index  of $\Gamma$ $\sim$ 2-3. Likewise, the {\it ASCA}
spectra of the northeast region of CTB 1 also show an excess at
higher energies: these spectra are fit best by a power law with a 
photon index $\Gamma$ = 1.4 plus the soft thermal component. The {\it Chandra}
observation of the southwestern region reveals localized regions of
hard emission: one such region is $\sim 1'$ in size. The X-ray 
spectrum of this region may be fit with either a higher temperature
thermal component ($\it{kT}$ = 0.66 keV) or with the combination of a softer
thermal component ($\it{kT}$ = 0.28 keV) and a power law component 
($\Gamma$ $\sim$ 2.0). Because of the poor photon statistics, 
its true nature is unclear. Possible scenarios for its origin include
temperature variations within the X-ray emitting plasma of CTB 1, including
the ejecta, or localized non-thermal X-ray emission. 
%Due to the limited number of counts associated
%with this emission: its true nature is still unclear. Plausible scenarios
%for its origin include i) temperature variations across CTB 1, 
%ii) ejecta emission characterized by a higher temperature or iii) 
%localized non-thermal X-ray emission. 

%We found CTB1 is ejecta dominated yet old SNR..
%While it is also possible for W44 has abduance gradient or differences region by region
%because ejecta are highly structureed, at this point it is unclear 
%it is due to real ejecta contirbution or due to lack of accounting the dust destrucction.

4. The {\it ASCA} hard ($E > 1$ keV) image of CTB 1 reveals a point-like 
source seen in projection against
the diffuse emission of CTB 1. This source -- denoted as 1WGA J0001.4$+$6229
and located at RA (J2000.0) 00$^h$ 01$^m$ 25.$^s$5, Dec (J2000.0) 
$+$62$^{\circ}$ 29$\arcmin$ 40$\arcsec$ -- may be
a neutron star associated with CTB 1. The
GIS2/GIS3 spectra of this source are well-fit by a power-law continuum
with a photon index $\Gamma$=2.2$^{+0.5}_{-1.2}$ (typical for rotation-powered 
pulsars) and the measured column density is comparable to the column density measured
for CTB 1. There is marginal evidence for pulsations in X-ray data at
47.6 msec, but no pulsations have been detected at radio wavelengths. 

\acknowledgments

We thank the referee for many useful comments which helped improve the
overall quality of the manuscript.
We acknowledge useful discussions with Steven Reynolds regarding the
nature of the hard X-ray emission seen toward CTB 1. T.G.P. thanks
Ken Ebisawa and Koji Mukai for their assistance with analyzing the
{\it ASCA} data, Keith Arnaud for suggestions during the spectral
fitting process and Ilana Harrus for her assistance with making the
mosaicked X-ray images of HB21 and CTB 1. T.G.P. also thanks Daniel Harris,
Samantha Stevenson and Nicholas Lee for helpful suggestions regarding the
reduction of the {\it Chandra}
observations of CTB 1. We also thank Robert Fesen for kindly
sharing his optical images of CTB 1 with us, Bryan Jacoby for
assistance with the GBT observations and Eric Gotthelf for
his contributions in searching for pulsed X-ray emission from 
1WGA J00001.4$+$6229. 
This research has made use of NASA's Astrophysics 
Data System and data obtained through the High Energy 
Astrophysics Science Archive Research Center Online Service, provided by 
the NASA/Goddard Space Flight Center. The research presented in this
paper has used data from the Canadian Galactic Plane Survey, a 
Canadian project with international partners, supported by the
Natural Sciences and Engineering Resources Council.

\clearpage

%% You can append references to a table using the \tablerefs command.
\begin{deluxetable}{lccccccccccccc}
\tabletypesize{\scriptsize}
\rotate
\tablecaption{Summary of {\it ASCA} GIS and SIS Observations of HB21
and CTB 1\label{ASCAObsTable}}
\tablewidth{0pt}
\tablehead{
& & & & & & \colhead{GIS2} & \colhead{GIS2} & \colhead{GIS3} &
\colhead{GIS3}
& \colhead{SIS0} & \colhead{SIS0} & \colhead{SIS1} & \colhead{SIS1} \\
& & & & & & \colhead{Effective} & \colhead{Count} & \colhead{Effective}
& \colhead{Count} & \colhead{Effective} & \colhead{Count} &
\colhead{Effective} & \colhead{Count} \\
& & & & \colhead{Right} & & \colhead{Exposure} & \colhead{Rate}
& \colhead{Exposure} & \colhead{Rate} & \colhead{Exposure} &
\colhead{Rate}
& \colhead{Exposure} & \colhead{Rate}\\
\colhead{Sequence} & & \colhead{} & \colhead{Observation}
& \colhead{Ascension} & \colhead{Declination} & \colhead{Time} &
\colhead{(10$^{-2}$ cts} & \colhead{Time} & \colhead{(10$^{-2}$ cts}
& \colhead{Time} & \colhead{(10$^{-2}$ cts} & \colhead{Time}
& \colhead{(10$^{-2}$ cts} \\
\colhead{Number} & \colhead{SNR} & \colhead{Pointing} & \colhead{Date} &
\colhead{(J2000.0)} & \colhead{(J2000.0)} & \colhead{(s)} &
\colhead{sec$^{-1}$)} & \colhead{(sec)} & \colhead{s$^{-1}$)}
& \colhead{(s)} & \colhead{s$^{-1}$)} & \colhead{(s)} &
\colhead{s$^{-1}$)}
}
\startdata
55053000 & HB21 & NW & 1997 June 9-10 & 20 44 53.8 & $+$50 54 31 &
39254 & 7.38$\pm$0.19 & 39248 & 7.31$\pm$0.19 & 35307 & 6.78$\pm$0.18 &
35158 & 5.55$\pm$0.17\\
55054000 & HB21 & SE & 1997 June 10-11 & 20 46 31.4 & $+$50 38 43 &
38415 & 7.18$\pm$0.19 & 38413 & 7.39$\pm$0.19 & 34448 & 5.52$\pm$0.17 &
33940 & 4.21$\pm$0.16 \\
54026000 & CTB1 & SW & 1996 January 21 & 23 57 31.9 & $+$62 25 20 &
58065 & 6.03$\pm$0.15 & 58078 & 5.44$\pm$0.14 & 53592 & 10.52$\pm$0.15 &
53746 & 8.64$\pm$0.13\\
54027000 & CTB1 & NE & 1996 January 22 & 00 01 21.9 & $+$62 38 52 &
41637 & 4.83$\pm$0.16 & 41649 & 5.16$\pm$0.17 & \nodata & \nodata &
\nodata & \nodata
\enddata
\tablecomments{The units of Right Ascension are hours, minutes and
seconds and the units of Declination are degrees, arcminutes and
arcseconds. Count rates are for the energy range 0.6--10.0 keV.}
\end{deluxetable}

\clearpage

\begin{deluxetable}{lcccccc}
\rotate
\tabletypesize{\scriptsize}
\tablecaption{Summary of Fits to GIS and SIS Spectra of 
HB21\tablenotemark{a}\label{HB21SpectralTable}}
\tablewidth{0pt}
\tablehead{
\colhead{Parameter} & \colhead{GIS2/3+SIS0/1} & \colhead{GIS2/3+SIS0/1} 
& \colhead{GIS2/3+SIS0/1} & \colhead{GIS2/3+SIS0/1} & \colhead{GIS2} & \colhead{GIS2}
}
\startdata
Region & Northwest & Northwest & Southeast & Southeast & Northwest & Northwest \\
& & & & & and Southeast & and Southeast \\
Model\tablenotemark{b} & PHABS$\times$ & PHABS$\times$ & PHABS$\times$ & 
PHABS$\times$ & PHABS$\times$ & PHABS$\times$\\
& VAPEC  & VNEI & VAPEC & VNEI & VAPEC & VNEI \\
$\chi$$^2_{\nu}$ ($\chi$$^2$/DOF) & 1.05 (1103/1054) & 1.05 (1102/1053) 
& 1.06 (1119/1054) & 1.04 (1099/1053) & 1.06 (340.43/321) & 1.05 (336.50/327) \\ 
$N$$_H$ ($\times$ 10$^{22}$ cm$^{-2}$) & 0.23$\pm$0.04 & 0.24$^{+0.06}_{-0.04}$ 
& 0.31$\pm$0.04 & 0.30$\pm$0.06 & 0.22$^{+0.05}_{-0.06}$ & 0.24$\pm$0.06 \\
${kT}$ (keV) & 0.65$\pm$0.03 & 0.63$^{+0.03}_{-0.04}$ & 0.68$^{+0.03}_{-0.02}$ 
& 0.67$\pm$0.04 & 0.63$\pm$0.06 & 0.62$^{+0.06}_{-0.05}$ \\
Si & 1.3$^{+0.3}_{-0.2}$ & 1.8$\pm$0.5 & 1.4$\pm$0.3 & 2.0$\pm$0.4 & 1.6$^{+0.5}_{-0.3}$
& 1.9$^{+0.5}_{-0.4}$ \\
S & 2.4$\pm$1.0 & 3.6$^{+1.7}_{-1.9}$ & 1.7$^{+0.9}_{-0.8}$ & 3.0$\pm$1.4 &
3.4$^{+1.6}_{-1.4}$ & 4.3$^{+1.9}_{-1.8}$ \\
$\tau$ (10$^{11}$ cm$^{-3}$ s) & \nodata & 5.9($>$3.2) & \nodata & 
4.1$^{+5.9}_{-1.1}$ & \nodata & 350($>$4.0) \\
EM\tablenotemark{c} /(4$\pi$d$^2$/10$^{-14}$) 
(cm$^{-5}$) & 4.4$\times$10$^{-3}$ & 4.4$\times$10$^{-3}$ & 
4.7$\times$10$^{-3}$ & 4.4$\times$10$^{-3}$ & 4.0$\times$10$^{-3}$ &
4.5$\times$10$^{-3}$  \\
Absorbed Flux\tablenotemark{d} (ergs cm$^{-2}$ s$^{-1}$) &
5.2$\times$10$^{-12}$ & 5.1$\times$10$^{-12}$ & 
4.7$\times$10$^{-12}$ & 4.6$\times$10$^{-12}$ & 5.1$\times$10$^{-12}$ &
5.0$\times$10$^{-12}$ \\
Unabsorbed Flux\tablenotemark{d} (ergs cm$^{-2}$ s$^{-1}$) &
1.1$\times$10$^{-11}$ & 1.1$\times$10$^{-11}$ & 1.1$\times$10$^{-11}$
& 1.1$\times$10$^{-11}$ & 9.8$\times$10$^{-12}$ & 1.1$\times$10$^{-11}$  \\
Unabsorbed Luminosity\tablenotemark{d} (ergs s$^{-1}$) & 3.8$\times$10$^{33}$ 
& 3.8$\times$10$^{33}$ & 3.8$\times$10$^{33}$ & 3.8$\times$10$^{33}$ 
& 3.4$\times$10$^{33}$ & 3.7$\times$10$^{33}$ \\
\enddata
\tablenotetext{a}{All quoted error bounds are 90\% confidence intervals.}
\tablenotetext{b}{PHABS is a photoelectric absorption model, VAPEC
is a thermal plasma model in ionization equilibrium, and  VNEI
is a nonequilibrium ionization thermal model (see Section \ref{HB21Section} for references 
for these models).}
\tablenotetext{c}{Emission measure, defined here as $\int n_e n_H$ 
dV: here, $d$ is the distance to HB21 (in cm) and $n$$_e$ and $n$$_H$ are
the electron and H densities (in cm$^{-3}$).}
\tablenotetext{d}{For the energy range 0.6--10.0 keV. The luminosity 
estimates are for an assumed distance of 1.7 kpc.}
\end{deluxetable}

\clearpage
\begin{deluxetable}{lcccccc}
\rotate
\tabletypesize{\scriptsize}
\tablecaption{Summary of Fits to {\it Chandra} Spectra of 
CTB 1\tablenotemark{a}\label{ctb1chandrafit}}
\tablewidth{0pt}
\tablehead{
\colhead{Parameter} & \colhead{Diffuse Region} & \colhead{Soft Region} 
& \colhead{Hard Region} & \colhead{Hard Region} & \colhead{Hard Region}
}
\startdata
Model & PHABS$\times$ & PHABS$\times$ & {\bf PHABS$\times$} & PHABS$\times$ & 
PHABS$\times$ & \\
& VAPEC & VAPEC & {\bf  VAPEC} & POWER LAW & (APEC+POWER LAW)  & \\ 
$\chi$$^2_{\nu}$ ($\chi$$^2$/DOF) & 0.99 (169.72/172) 
& 0.94 (122.77/131) & 0.41$^d$ [0.41]$^e$  & 0.50$^d$   
& 0.40$^d$ &\\ 
$N$$_H$ ($\times$ 10$^{22}$ cm$^{-2}$) & 0.64$\pm$0.08
& 0.56$^{+0.11}_{-0.16}$ & 0.18($<$0.75) [0.36$^{+0.39}_{-0.25}$] & 
0.76($^{+0.64}_{-0.57}$) & 0.47(--) &  \\
${kT}$$_1$ (keV) & 0.28$\pm$0.03 & 0.28$^{+0.06}_{-0.04}$ & 
0.66$^{+0.27}_{-0.41}$ [0.63$^{+0.17}_{-0.53}$] & \nodata & 0.28 (frozen) & \\
O & 1.7$^{+1.2}_{-0.6}$ & 1.8$^{+2.2}_{-0.8}$ &  1.7 [$\equiv$1] & \nodata & 
1 (frozen)  &  \\
Ne & 1.6$^{+0.8}_{-0.5}$ & 1.1$^{+1.0}_{-0.4}$ & 1.1$^{+1.0}_{-0.4}$ [$\equiv$1] & 
\nodata & 1 (frozen) &  \\
Fe & 0.4$\pm$0.2 & 0.7$^{+0.6}_{-0.3}$ & 0.5$^{+3.0}_{-0.3}$ [$\equiv$1] & \nodata & 1 
(frozen) &  \\
EM$_1$\tablenotemark{b} /(4$\pi$d$^2$/10$^{-14}$) (cm$^{-5}$) & 
4.33$\times$10$^{-3}$ & 2.23$\times$10$^{-3}$ & 2.14$\times$10$^{-4}$  
& \nodata & 2.41$\times$10$^{-4}$  & \\ 
$\Gamma$ & \nodata & \nodata & \nodata & 6.4($^{+5.6}_{-2.9}$) & 1.9(--) &  \\
Normalization & \nodata & \nodata & \nodata & 2.08$\times$10$^{-4}$ & 
1.27$\times$10$^{-5}$ & \\
\hline
Counts\tablenotemark{c} & 9232 & 5319 & 979 & 979 & 979 &  \\
Absorbed Flux\tablenotemark{c} (ergs cm$^{-2}$ s$^{-1}$) &
5.29$\times$10$^{-13}$ & 3.62$\times$10$^{-13}$ & 
5.90$\times$10$^{-14}$ & 6.50$\times$10$^{-14}$ & 8.21$\times$10$^{-14}$
& \\
Unabsorbed Flux\tablenotemark{c} (ergs cm$^{-2}$ s$^{-1}$) &
8.66$\times$10$^{-12}$ & 4.70$\times$10$^{-12}$ & 
7.52$\times$10$^{-13}$ & 1.60$\times$10$^{-12}$ & 4.34$\times$10$^{-13}$ 
& \\
Luminosity\tablenotemark{c} (ergs s$^{-1}$) & 
9.96$\times$10$^{33}$ & 5.41$\times$10$^{33}$ & 8.65$\times$10$^{32}$
& 1.84$\times$10$^{33}$ & 4.99$\times$10$^{32}$ &  
\enddata
\tablenotetext{a}{All quoted error bounds are 90\% confidence intervals.
The best model is marked in bold where a few models 
are presented for the same spectra set.}
\tablenotetext{b}
{Defined as $\int n_e n_H$ dV: here, $d$ is the distance to CTB 1 
(in cm) and $n$$_e$ and $n$$_H$ are the electron and H densities 
(in cm$^{-3}$). 
%Expressed here as $\times$ 4$\pi$d$^2$/10$^{-14}$ D cm$^{-5}$.
}
\tablenotetext{c}{For the energy range 0.5--5.0 keV.}
\tablenotetext{d}{The numbers of degrees of freedom range from 101 to 104.}
\tablenotetext{e}{The numbers in brackets are  the fits with fixed solar abundnaces.}
\end{deluxetable}

\clearpage
\begin{deluxetable}{lccccccc}
\rotate
\tabletypesize{\scriptsize}
\tablecaption{Summary of Fits to GIS and SIS Spectra of
Southwest and Northeast Regions of CTB 1\tablenotemark{a}\label{ctb1ascafit}}
\tablewidth{0pt}
\tablehead{
\colhead{Parameter} & \colhead{GIS2/3+SIS0/1} & \colhead{GIS2/3+SIS0/1} &
\colhead{GIS2/3+SIS0/1} & \colhead{GIS2/3+SIS0/1} &
\colhead{GIS2/3+SIS0/1} & \colhead{GIS2/3+SIS0/1} & \colhead{GIS2/3}
}
\startdata
Region & Southwest & Southwest & Southwest & Southwest & Southwest &
Southwest & Northeast \\
Model & PHABS$\times$ & PHABS$\times$ & PHABS$\times$ & {\bf PHABS$\times$}
& PHABS$\times$ & {\bf PHABS$\times$} & PHABS$\times$\\
      & VAPEC & VNEI & (VAPEC+ & {\bf (VNEI+ }& (VAPEC+ & {\bf (VNEI+} & (VAPEC+\\
      &       &      & POWER LAW) & {\bf POWER LAW)} & VAPEC) & {\bf VNEI)} & POWER LAW)\\
$\chi$$^2$/DOF  & 1.33$^g$ [1.33]$^h$ & 1.29  & 1.15 [1.15]$^g$  
& 1.12$^g$ [1.19]$^h$  & 1.13$^g$  & 1.12$^g$ [1.12]$^h$ & 1.19$^i$ [1.16]$^h$  \\
%($\chi$$^2$/DOF) & (1378.92/1039) &  (1340.27/1038) & (1185.40/1036) &
%(1160.28/1033) &  (1167.485/1034) & (1153.87/1032) & (451.30/394) \\
$N$$_H$ ($\times$ 10$^{22}$ cm$^{-2}$) & 0.59 [ 0.51] & 0.68 & 0.55$^{+0.13}_{-0.15}$ 
& 0.57$^{+0.48}_{-0.19}$ [0.59] & 0.66$^{+0.09}_{-0.13}$ & 0.80$^{+0.3}_{-0.1}$ [0.9]
& 0.47$\pm$0.32 [0.60] \\
${kT}$$_{soft}$ (keV) & 0.30 [0.30] & 0.41 & 0.28$^{+0.09}_{-0.04}$ & 
0.27$^{+0.23}_{-0.15}$ [0.27] & 0.25$^{+0.05}_{-0.03}$ 
& 0.23$^{+0.17}_{-0.06}$ [0.22] & 0.22$^{+0.34}_{-0.12}$ [0.24]\\
O\tablenotemark{b} & 1.7 [$\equiv$1] & 1.7 & 1.7 & 1.7  [$\equiv$1] & 1.7 & 1.7 [$\equiv$1] & $\equiv$1.7 [$\equiv$1]\\
Ne & 1.6 [1.2] & 1.6 & 1.6 & 2.7($>$2.0) [$\equiv$1] & 2.1$\pm$0.4 & 2.0$^{+0.5}_{-1.1}$ [1.8] &
$\equiv$1.6 [$\equiv$1]\\
Fe & 0.5 [0.4] & 0.4 & 0.4 & 0.8($>$0.4) [$\equiv$1] & 0.7$\pm$0.4 & 1.2$^{+3.8}_{-0.5}$ [1.2] &
$\equiv$0.4 [$\equiv$1] \\
$\tau$$_1$ (10$^{11}$ cm$^{-3}$ s) & \nodata & 2 & \nodata 
& 10 ($>$0.2) [15] & \nodata & 0.4($<$5) [0.4] & \nodata \\
EM$_1$\tablenotemark{c}/(4$\pi$d$^2$/10$^{-14}$) (cm$^{-5}$) &
2$\times$10$^{-2}$ & 2$\times$10$^{-2}$ & 2$\times$10$^{-2}$ & 
2$\times$10$^{-2}$ & 3$\times$10$^{-2}$ & 8$\times$10$^{-2}$ &
2$\times$10$^{-2}$ \\
\hline
$\Gamma$ or ${kT}$$_{hard}$ & \nodata & \nodata & 
2.2$^{+0.4}_{-1.1}$ &  3.0$^{+1.8}_{-1.3}$ [3.0]
& 3.3$^{+4.7}_{-1.5}$ & 3.1($>2$) [3] & 1.4$\pm$0.9\\
%${kT}$$_2$ & \nodata & \nodata & \nodata & \nodata & 3.3$^{+5.7}_{-1.3}$
%& 3.1$^{+7.0}_{-1.1}$\\
$\tau$$_2$ (10$^{11}$ cm$^{-3}$ s) & \nodata & \nodata & \nodata & \nodata
& \nodata & 0.7$^{+1.2}_{-0.4}$ [0.7] & \nodata \\
EM$_2$\tablenotemark{c}/(4$\pi$d$^2$/10$^{-14}$) (cm$^{-5}$) &
\nodata & \nodata & \nodata & 1$\times$10$^{-3}$ & 1$\times$10$^{-3}$ 
& 8$\times$10$^{-4}$ & \nodata \\
Normalization\tablenotemark{e} & \nodata & \nodata &
4$\times$10$^{-4}$ & \nodata & \nodata & \nodata & 2$\times$10$^{-4}$ \\
\hline
Absorbed Flux\tablenotemark{f} & 3.52$\times$10$^{-12}$ &
3.64$\times$10$^{-12}$ & 4.44$\times$10$^{-12}$ & 
4.31$\times$10$^{-12}$ & 4.38$\times$10$^{-12}$ & 4.43$\times$10$^{-12}$ &
4.37$\times$10$^{-12}$ \\
Unabsorbed Flux\tablenotemark{f} & 3.37$\times$10$^{-11}$ &
5.35$\times$10$^{-11}$ & 3.34$\times$10$^{-11}$ & 
3.50$\times$10$^{-11}$ & 4.88$\times$10$^{-11}$ & 1.10$\times$10$^{-10}$
& 3.02$\times$10$^{-11}$ \\
%Absorbed Luminosity\tablenotemark{e} & 4.05$\times$10$^{33}$ &
%4.19$\times$10$^{33}$ & 4.96$\times$10$^{33}$ & 4.95$\times$10$^{33}$
%& 5.04$\times$10$^{33}$ & 5.09$\times$10$^{33}$\\
Luminosity\tablenotemark{e} & 3.88$\times$10$^{34}$ &
6.16$\times$10$^{34}$ & 4.19$\times$10$^{34}$ &   
4.03$\times$10$^{34}$ & 5.61$\times$10$^{34}$ & 1.27$\times$10$^{35}$
& 3.47$\times$10$^{34}$
\enddata
\tablenotetext{a}{All quoted error bounds are 90\% confidence intervals and the best model 
 is marked in bold.}
\tablenotetext{b}{The abundance for oxygen was frozen to 1.7 and 1 [numbers in brackets] for all
fits. Since the errors are similar for both cases, only one set of errors are given.}
\tablenotetext{c}{Emission measure, defined as $\int n_e n_H$ dV:
here, $d$ is the distance to CTB 1 (in cm) and $n$$_e$
and $n$$_H$ are the electron and H densities (in cm$^{-3}$).}
\tablenotetext{d}{The units for the thermal component ${\it kT}$$_{hard}$
are keV.}
\tablenotetext{e}{Normalization for the power law component in units of
photons keV$^{-1}$ cm$^{-2}$ s$^{-1}$ at 1 keV.}
\tablenotetext{f}{For the energy range 0.6--10.0 keV. The units for
the flux are ergs cm$^{-2}$ s$^{-1}$ and the units for the luminosity
are ergs s$^{-1}$. The luminosity estimates are for an assumed 
distance of 3.1 kpc.}
\tablenotetext{g,i}{The degree of freedom for Southwestern spectra ($^g$) is $\sim$1039 
with both GIS and SIS data,
for Northeastern spectra$^i$ is 394 with only GIS data.                            }
\tablenotetext{h}{The numbers in brackets are the fits with fixed solar abundnaces.}

\end{deluxetable}

\clearpage

\begin{deluxetable}{lcc}
\tablecaption{Summary of the Physical Properties of HB21 and 
CTB 1\label{physical}\tablenotemark{a}}
\tablewidth{0pt}
\tablehead{
\colhead{Property} & \colhead{HB21} & \colhead{CTB 1}
} 
\startdata
Angular size (arcmin) & 120 $\times$ 90 & 34 \\
Distance (kpc) & 1.7 & 3.1 \\
Physical Diameter (pc) & 59$\times$45 & 31 \\
Average Radius (pc) & 26 & 15.3 \\
$N$$_H$ (10$^{21}$ cm$^{-2}$) & 0.23--0.31 & 0.57--0.80 \\
$kT$ or $\Gamma$ & 0.63--0.68 keV & 0.28 keV, $\Gamma$ = 2--3 \\
                 &                & (or 3 keV) \\
$[$O/Fe$]$ & -- & 4.3$^{+10.2}_{-2.5}$ \\
$[$Ne/Fe$]$ & -- & 4.0$^{+8.0}_{-2.2}$ \\
$n$$_e$ (cm$^{-3}$)  & 0.06 & 0.16$f$$^{-1/2}_{soft}$\\ 
$M$$_X$\tablenotemark{b} (M$_{\odot}$) & 23.4 & 40$f$$^{1/2}_{soft}$\\
X-ray pressure $P$/$k$(K cm$^{-3}$) & 0.9$\times$10$^6$ & 2.8$\times$10$^6$ \\
Shock velocity (optical)\tablenotemark{c} (km/s) & \nodata & 400\\
Age (optical)\tablenotemark{c} (yr) & \nodata & 1.6$\times$10$^4$ \\
Shock velocity (HI shell)\tablenotemark{d} (km/s) & 124 & 107$?$\\
Age (HI shell)\tablenotemark{d} (yr) & 4.5$\times$10$^4$ & 4.4$\times$10$^4$\\
$n$$_0$ (HI shell)\tablenotemark{d} (cm$^{-3}$) & 2.7 & \nodata \\
$E$$_0$ (HI shell)\tablenotemark{d} (ergs) & 1.6$\times$10$^{51}$ & \nodata \\
$\Delta$$M$ (HI shell)\tablenotemark{d} (M$_{\odot}$) & 150$\pm$10 & 12$\pm$8 
\enddata
\tablenotetext{a}{See Sections \ref{HB21SubSection} and \ref{CTB1EjectaSection}.}
\tablenotetext{b}{In the case of HB21, the listed X-ray emitting mass is for 
the volume of the whole SNR. The X-ray emitting
mass of just the area of HB21 observed by {\it ASCA} is 2.6 M$_{\odot}$.}
\tablenotetext{c}{\citet{Fesen97}.}
\tablenotetext{d}{\citet{Koo91}, who assumed distances of 1.1 kpc and 3.2 kpc to
HB21 and CTB 1, respectively.}
\end{deluxetable}

\clearpage

\clearpage

\clearpage

\clearpage

%\begin{figure}
%\plotone{ConfCon_HB21_NW_PHABSVAPEC_NHKT.ps}
%\caption{Confidence contours for $N$$_H$ and $kT$ for the PHABS*VAPEC
%fit to the extracted GIS2, GIS3, SIS0 and SIS1 spectra of the 
%northwestern portion of HB21.\label{fig5}}
%\end{figure}

\clearpage

%\begin{figure}
%\psfig{figure=ConfCon_HB21_NW_PHABSVAPEC_SIS.ps,height=6truecm,angle=270}
%%\plotone{ConfCon_HB21_NW_PHABSVAPEC_SIS.ps} 
%\caption{
%************************************************
%IT WOULD BE MORE INTERESTING TO SEE SI(S) VS. T. REMOVE UNNECESSARY TITLE, 
%FILE, AND DATE TEXTS.
%************************************************
%Confidence contours for silicon and sulfur based on thermal fits
%to the GIS2, GIS3, SIS0 and SIS1 spectra of the 
%northwestern portion of HB21.}
%\label{hb21abundance}
%\end{figure}

%\clearpage

\clearpage

\clearpage

%\clearpage

\clearpage

\clearpage

\clearpage

\clearpage

\newpage
\clearpage

\newpage

\clearpage

%begin{figure}
%\epsscale{1.0}
%\plotone{hb21finalascaspec1.ps}
%\caption{{\it ASCA} spectra of the northeastern portion of
%CTB 1.\label{fig12}}
%\end{figure}
\newpage
\clearpage

\end{document}